\documentclass{aastex}          
\usepackage{spr-astr-addons}    
\usepackage{graphicx}
\usepackage{wrapfig}
\usepackage{amsmath}
\usepackage{amssymb}
\usepackage{latexsym}
\usepackage{color}
\usepackage{subfigure}
\usepackage{hyperref}

\renewcommand{\vec}[1]{\boldsymbol{#1}}

\begin{document}
%
\title{Kelvin--Helmholtz instability in an active region jet observed with \emph{Hinode}}

\shorttitle{KH instability in an EUV solar jet}
\shortauthors{Zhelyazkov et al.}

\author{I.~Zhelyazkov\altaffilmark{1}}
\affil{Faculty of Physics, Sofia University, 1164 Sofia, Bulgaria}
\and
\author{R.~Chandra\altaffilmark{2}}
\affil{Department of Physics, DSB Campus, Kumaun University, Nainital 263\,002, India}
\and
\author{A.~K.~Srivastava\altaffilmark{3}}
\affil{Department of Physics, Indian Institute of Technology (Banaras Hindu University), Varanasi 221\,005, India}

\altaffiltext{1}{Faculty of Physics, Sofia University, 1164 Sofia, Bulgaria}
\altaffiltext{2}{Department of Physics, DSB Campus, Kumaun University, Nainital 263\,002, India}
\altaffiltext{3}{Department of Physics, Indian Institute of Technology (Banaras Hindu University), Varanasi 221\,005, India}

\begin{abstract}
Over past ten years a variety of jet-like phenomena were detected in the solar atmosphere, including plasma ejections over a range of coronal temperatures being observed as extreme ultraviolet (EUV) and X-ray jets.  We study the possibility for the development of Kelvin--Helmholtz (KH) instability of transverse magnetohydrodynamic (MHD) waves traveling along an EUV jet situated on the west side of NOAA AR 10938 and observed by three instruments on board \emph{Hinode\/} on 2007 January 15/16 (Chifor et al., Astron.\ Astrophys.\ \textbf{481}, L57 (2008)).  The jet was observed around $\mathrm{Log}\, T_\mathrm{e} = 6.2$ with up-flow velocities exceeded $150$~km\,s$^{-1}$.  Using Fe {\footnotesize\textsc{XII}} $\lambda186$ and $\lambda195$ line ratios, the measured densities were found to be above $\mathrm{Log}\, N_\mathrm{e} = 11$.  We have modeled that EUV jet as a vertically moving magnetic flux tube (untwisted and weakly twisted) and have studied the propagation characteristics of the kink ($m = 1$) mode and the higher $m$ modes with azimuthal mode numbers $m = 2, 3, 4$.  It turns out that all these MHD waves can become unstable at flow velocities in the range of $112$--$114.8$~km\,s$^{-1}$.  The lowest critical jet velocity of $112$~km\,s$^{-1}$ is obtained when modeling the jet as compressible plasma contained in an untwisted magnetic flux tube.  When the jet and its environments are treated as incompressible media, the critical jet velocity becomes higher, namely $114.8$~km\,s$^{-1}$.  A weak twist of the equilibrium magnetic field in the same approximation of incompressible plasmas slightly decreases the threshold Alfv\'en Mach number, $M_{\mathrm{A}}^{\mathrm{cr}}$, and consequently the corresponding critical velocities, notably to $114.4$~km\,s$^{-1}$ for the kink mode and to $112.4$~km\,s$^{-1}$ for the higher $m$ modes.  We have also compared two analytically found criteria for predicting the threshold Alfv\'en Mach number for the onset of KH instability and have concluded that one of them yields reliable values for $M_{\mathrm{A}}^{\mathrm{cr}}$.  Our study of the nature of stable and unstable MHD modes propagating on the jet shows that in a stable regime all the modes are pure surface waves, while the unstable kink ($m = 1$) mode in untwisted compressible plasma flux tube becomes a leaky wave.  In the limit of incompressible media (for the jet and its environment) all unstable modes are non-leaky surface waves.
\end{abstract}


\section{Introduction}
\label{sec:intro}
Solar jets are ubiquitous plasma ejecta in the Sun's atmosphere and are extensively observed in H$\alpha$ \citep{roy}, Ca {\footnotesize\textsc{II}} H \citep{shi07,chi08a,nis}, EUV \citep{ale}, and soft X-ray \citep{shi92} indicating their formation by the multi-temperature plasma \citep{chi08a}. Extreme ultraviolet (EUV) coronal jets are either straight or curved, as well as inverted Y-shaped plasma ejecta observed mostly in the transition region and  corona at different temperatures.  They may appear in the imaging observations as bright, dark, or dark-bright plasma structures as an emission or absorption features in various coronal/TR wavelengths \citep{chae,chen,liu,ko,jia,nis}.  The intra-relationship of the jets at different emissions has been explored in great details, e.g., in the H$\alpha$ and soft X-ray \citep{rust,bri,canf}, H$\alpha$ and EUV \citep{schm,chae,jia,chen}, and EUV and soft X-ray \citep{ale,kim,chi08a} emissions, which provides the clues on their exact driving mechanisms.  The cool and hot component of the typical coronal jets have been physically explained by various magnetic reconnection based numerical models \citep{yoko95,yoko96}.

\cite{kim} have investigated in greater details the three AR jets simultaneously observed by \emph{Hinode\/} X-ray telescope and the \emph{Transition Region and Coronal Explorer\/} (\emph{TRACE}; \citealp{handy}) and demonstrated that the EUV and soft X-ray components of the coronal jet have similar ballistic speeds, life-time and spatial size.  \cite{kim} have observed an active-region jet using various spectral lines formed at different temperatures as observed by \emph{Hinode}/EIS.  They have found simultaneous blue-shift (upflows) up to maximum speed of $-64$~km\,s$^{-1}$ and red-shift (downflows) up to $20$~km\,s$^{-1}$ at the base of the coronal jet as observed by \emph{Hinode}/EIS. Moreover, \cite{chi08b} have observed a blue-shift of the active region jet plasma with the apparent velocity of $150$~km\,s$^{-1}$ and comparatively a weak red-shifted plasma motion at the base of the jet.  \cite{chi08a} have reported that the EUV jet was co-spatial and consists of the similar kinematical features as of its soft X-ray counterpart.  A similar conclusion has been drawn by \cite{yang} who have presented simultaneous observations of three recurring jets in EUV and soft X-ray emissions that occurred in an active region on 2007 June 5.  On comparing their morphological and kinematic characteristics in these two different wavelengths and related emissions, they found that EUV and soft X-ray jets have co-spatial triggering site, similar direction, size, and propulsion speeds.  \cite{yang} have also analyzed jet's spectral properties by using multiple spectral lines as observed by \emph{Hinode}/EIS. They have found that these jets have temperature in the range of $0.05$ to $2.0$~MK, while maximum electron densities have been reported as $6.6 \times 10^9$ to $3.4 \times 10^{10}$~cm$^{-3}$.  For each of these observed jets, a blue-shifted component of the plasma and a red-shifted base are simultaneously observed.  These observed jets as reported by \cite{yang} have maximum Doppler velocities ranging from $25$ to $121$~km\,s$^{-1}$ for the blue-shifted plasma component, while it is observed in the range of $115$ to $232$~km\,s$^{-1}$ for the red-shifted plasma component.  These observational results are found to be in agreement with the magnetic reconnection driven coronal jet models owing to the direct $\vec{j} \times \vec{B}$ force \citep{yoko95,yoko96}.

Recently, \cite{chan} have presented and discussed the multi-wavelength observations of five homologous recurrent EUV solar jets triggered in NOAA AR 11133 on December 11, 2010, which were observed by the Atmospheric Imaging Assembly (AIA; \citealp{lem}) onboard the \emph{Solar Dynamic observatory\/} (\emph{SDO}; \citealp{pes}). The speed of these jets lie between $86$ and $267$~km\,s$^{-1}$, and thhey are triggered in the direction of open field lines as shown by Potential-field Source-surface (PFSS) extrapolations \citep{scha,alts,wang}.  \cite{cheu} have presented the evolution of four homologous helical recurrent jets triggered from NOAA AR 11793 in the transition region and corona on July 21, 2013 by using the AIA/\emph{SDO\/}. These jets have also been observed by the \emph{Interface Region Imaging Spectrograph\/} (IRIS; \citealp{bart}) and the Doppler velocities in these jets have been estimated using the spectral data.  They found that one edge of each of these jets was blue-shifted while the opposite edge was red-shifted simultaneously, which indicate that these jets have helical structure with the same sign of the helicity.

Magnetically structured EUV coronal jets that we consider as moving cylindrical magnetic flux tubes likewise other solar flowing structures support the propagation of various kind of magnetohydrodynamic (MHD) waves and oscillations.  In the case when the tube plasma density is radially inhomogeneous along with the basic MHD waves, e.g., slow magnetoacoustic, fast magnetoacoustic (kink and sausage), and torsional Alfv\'en waves, there appear so called continuous slow and Alfv\'en spectra in the frequency domain where discrete global modes exist \citep{appe}.  Rotation, sheared flow velocities and/or ambient magnetic fields as well as resistivity make the wave spectra much more complicated and difficult for analytical treatment.  An excellent explanation of all the variety of MHD oscillations and waves in bounded cylindrical plasmas the reader can find in the books by \cite{goed04} and \cite{goed10}. A useful review on the nonlinear MHD waves in the solar atmosphere and more specifically in the magnetically structured flux tubes was presented by \cite{rude}.

On contrary to the case when the solar atmospheric plasma is static and propagating modes are stable, the axial motion of flux tubes leads to a velocity jump at the tube surface that can result further into the evolution of a special kind of instability known as Kelvin--Helmholtz (KH) instability.  The KH instability arises at the interface of two fluid layers that move with different speeds. Flows are general uniform in both the layers, however, the strong velocity shear arises near the interface region of these two fluids forming a vortex sheet that becomes unstable to the spiral-like perturbations \citep{zaq15}.  When a magnetic flux tube moves along its axis then a vortex sheet is evolved near its boundary which may become unstable subjecting to the KH instability in a case when its axial speed exceeds some critical threshold value. This vortex sheet causes the conversion of the directed flow energy into the turbulent energy (e.g., \citealp{masl}).

The interest in exploring the KH instability in the solar atmosphere over the past decade arose from the circumstances that KH vortices were observed in solar prominences \citep{berg,rita,mart}, in Sweet--Parker current sheets \citep{lour}, in a coronal streamer \citep{feng}, and at boundaries of rising coronal mass ejections \citep{ofman,foul2011,foul2013,mostl}, inspite of their existence in variety of astrophysical and geophysical plasmas also \citep{bond,casa,pu,foul2010,chun,min,zhan,kepp99,miur}.  All these observations stimulated the modeling of KH instability in moving magnetic flux tubes.  \cite{zaq10} have studied the KH instability in twisted flux tubes moving in non-magnetic environment, \cite{sol} in magnetic tubes of partially ionized plasma, \cite{zaq11}, \cite{zhel12a}, and \cite{aja} in spicules, \cite{vash} and \cite{zhel12b} in soft X-ray jets, \cite{zhel} in photospheric tubes, \cite{zhel15a,zhel15b} in high-temperature and cool surges, \cite{zhel15c} in dark mottles, \cite{mostl} and \cite{zhel15d} at the boundary of rising CMEs, and \cite{zaq15} in rotating, tornado-like magnetized jets.  A review on modeling the KH instability in solar atmosphere jets (primarily in the limits of homogeneous ideal plasma) the reader can see in \citealp{zhel15e} and references therein.

In this paper, we study the KH instability in an active region (AR) EUV jet observed with \emph{Hinode\/} and detected by  \cite{chi08b} in the approximation of compressible plasma (untwisted flux tube) and incompressible plasma (twisted flux tube).  In the next section we specify the physical parameters of the jet, the topology of the magnetic fields inside and outside the moving flux tube modeling the jet and also discuss the wave dispersion equations for both geometries (untwisted and twisted tubes, respectively).  Section 3 deals with the numerical solving MHD wave dispersion relations, and in Sect.~4 we discuss the conditions under which the KH instability can develop in such moving structures.  In the last section we summarize the main results obtained in this article.

\section{Basic jet parameters, geometry, and MHD wave dispersion relations}
\label{sec:basic}
\cite{chi08b} have presented a study of an AR jet observed by the EUV Imaging Spectrometer (EIS) on board \emph{Hinode\/} on 2007 January 15/16 (west of NOAA AR 10938).  EIS covers two wavelength bands: $170$--$211$~\AA\ and $246$--$292$~\AA\, referred to as the short wavelength (SW) and long wavelength (LW) bands, respectively.  The jet was observed at temperatures between $5.4$ and $6.4$ in Log$\,T_\mathrm{e}$.  In the Ca {\footnotesize\textsc{XVII}} $\lambda192$ window at the location of the up-flow jet component, the O {\footnotesize\textsc{V}} $\lambda192.90$ line was seen next to the $\lambda192.83$ line (believed to be Fe {\footnotesize\textsc{XI}} according to \cite{young}).  \cite{chi08b} have observationally captured that there was also O {\footnotesize\textsc{V}} emissions near the base (the red-shifted plasma component) of the jet, but that could be due to the red-shifted Fe {\footnotesize\textsc{XI}} line.  A strong blue-shifted component and the signature of a weak red-shifted plasma component at the base of the jet was observed around $\mathrm{Log}\,T_\mathrm{e} = 6.2$ by \cite{chi08b}.  The up-flow velocities exceeded $150$~km\,s$^{-1}$, which is very significant observation.  The jet plasma was seen over a wide range of the temperatures between $5.4$ and $6.4$ in $\mathrm{Log}\,T_\mathrm{e}$.  Using Fe {\footnotesize\textsc{XII}} $\lambda186$ and $\lambda195$ line ratios, \cite{chi08b} have measured the electron densities above $\mathrm{Log}\,N_\mathrm{e} = 11$ for the high-velocity up-flowing plasma component of the observed coronal jet.

Our choice of jet's physical parameters are: $\mathrm{Log}\,T_\mathrm{i} = 6.08$ that corresponds to $T_\mathrm{i} = 1.2$~MK (subscript label `i' stands for \emph{interior}), and $\mathrm{Log}\,n_\mathrm{i} = 11$ that yields electron number density $n_\mathrm{i} = 1.0 \times 10^{11}$~cm$^{-3}$.  Bearing in mind that this EUV jet is positioned in the TR/lower corona (see Fig.~53(b) in \citealp{cheung}), we assume that AR plasma has a typical temperature $T_\mathrm{e} = 2$~MK (subscript label `e' stands for \emph{exterior}) and ambient coronal density $\rho_\mathrm{e} = 1.0 \times 10^{-13}$~g\,cm$^{-3}$ \citep{kur}, or equivalently an electron number density $n_\mathrm{e} = 5.98 \times 10^{10}$~cm$^{-3}$.  For simplicity we will treat the jet and its environment as ideal plasmas and will consider the densities in both media as homogeneous ones.  Moreover, we will neglect the influence of gravity on the wave propagation characteristics because the local gravitational scale length, $H \equiv c_\mathrm{s}^2/\gamma g$ ($c_\mathrm{s}$ being the typical sound speed, $\gamma = 5/3$ the adiabatic index, and $g = 273.95$~m\,s$^{-2}$ the local gravitational acceleration \citep{eric}), in our case equal to $40$--$60$~Mm, is larger than both the height of the jet and the wavelengths of the waves propagating along it.  Our choice for the background homogeneous magnetic field in TR/lower corona is $B_\mathrm{e} = 7$~G \citep{chae03}.  With a density contrast $\eta = \rho_\mathrm{e}/\rho_\mathrm{i} = 0.598$, the sound and Alfv\'en speeds in both media are: $c_\mathrm{si} \cong 128$~km\,s$^{-1}$, $v_\mathrm{Ai} = 47.6$~km\,s$^{-1}$ (more exactly $v_\mathrm{Ai} = 47.567$~km\,s$^{-1}$), and $c_\mathrm{se} = 166$~km\,s$^{-1}$, $v_\mathrm{Ae} \cong 62$~km\,s$^{-1}$, respectively.  We note that the magnetic field inside the jet is $B_\mathrm{i} = 6.899$~G that yields a ratio of both magnetic fields $b = B_\mathrm{e}/B_\mathrm{i} = 1.014$.  The plasma betas in the jet and its environment are correspondingly $\beta_\mathrm{i} = 8.76$ and $\beta_\mathrm{e} = 8.48$.  Thus, we can, in principle, consider both media as incompressible plasmas.
\begin{figure}[t]
   \centering
   \includegraphics[height=.30\textheight]{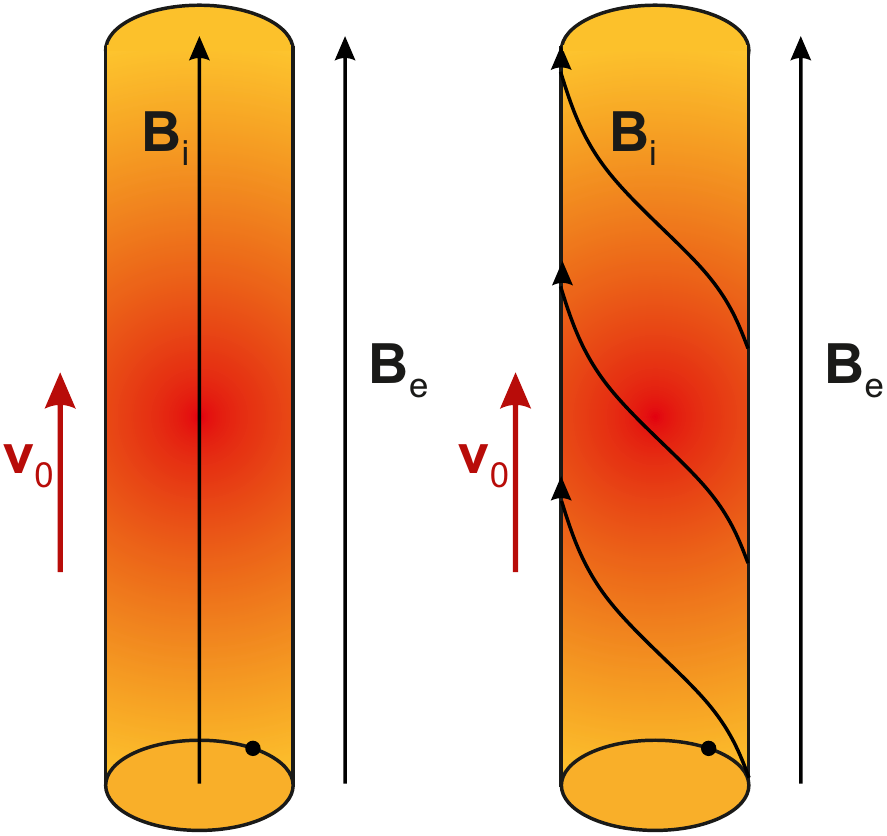}
   \caption{Equilibrium magnetic field geometries of an EUV solar jet in an untwisted flux tube (left picture) and in a weakly twisted flux tube (right picture).}
   \label{fig:fig1}
\end{figure}

As we already said, the EUV jet is modeled as a moving flux tube in two magnetic configuration: (i) as an untwisted tube with radius $a$ embedded in a homogeneous magnetic field $\vec{B}_\mathrm{i}$ and surrounded by coronal plasma immersed in the homogeneous magnetic field $\vec{B}_\mathrm{e}$ (see the left column in Fig.~\ref{fig:fig1}), and (ii) as a weakly twisted flux tube with a homogeneous magnetic field environment $\vec{B}_\mathrm{e}$ (see the right column in Fig.~\ref{fig:fig1}).  Our frame of reference is attached to the jet's environment, therefore the velocity of the moving magnetic flux tube, $\vec{v}_0$, has the meaning of a relative velocity if there is any ambient flow.  In our cylindrical coordinate system $(r, \phi, z)$, the twisted magnetic field $\vec{B}_\mathrm{i} = \left(0, B_{\mathrm{i}\phi}(r), B_{\mathrm{i}z}(r)\right)$ is characterized by the parameter $\varepsilon = B_{\mathrm{i}\phi}/B_{\mathrm{i}z}$, where both components are evaluated at the inner surface of the tube (in the trivial case of untwisted tube we simply have $\vec{B}_\mathrm{i} = (0,0,B_\mathrm{i})$ and, hence, $\varepsilon = 0$).  In cylindrical equilibrium, the magnetic field $\vec{B}_\mathrm{i}$ and thermal pressure $p_{\rm i}$ satisfy the equilibrium condition in the radial direction
\begin{equation}
\label{eq:equilib}
    \frac{\mathrm{d}}{\mathrm{d}r}\left( p_\mathrm{i} + \frac{B_\mathrm{i}^2}{2\mu}
    \right) = -\frac{B_{\mathrm{i} \phi}^2}{\mu r},
\end{equation}
where, $B_\mathrm{i}(r) = ( B_{\mathrm{i} \phi}^2 + B_{\mathrm{i} z}^2 )^{1/2}$ denotes the strength of the equilibrium magnetic field, and $\mu$ is the magnetic permeability.  We note that in Eq.~(\ref{eq:equilib}) the total (thermal plus magnetic) pressure gradient is balanced by the tension force (the right-hand side of Eq.~(\ref{eq:equilib})) owing to the twisted field.  Further on, in studying MHD wave propagation on the moving twisted tube, we will treat the jet's plasma as incompressible medium and will assume a magnetic fields' equilibrium with uniform twist, that is, the one for which $B_{\mathrm{i} \phi}(r)/r B_{\mathrm{i} z}(r)$ is a constant.  In such a case, the magnetic field inside the tube is $\vec{B}_\mathrm{i} = (0, Ar, B_{\mathrm{i} z})$, where $A$ and $B_{\mathrm{i} z}$ are constant.  For a magnetic flux tube with uniform twist the parameter $\varepsilon$ has the simple form of $Aa/B_{\mathrm{i} z}$, that is, it is a constant.  We note also that the aforementioned parameter $b$ (the ratio of axial magnetic fields) for a uniformly twisted flux tube is obviously $b_\mathrm{twist} = B_\mathrm{e}/B_{\mathrm{i} z}$.  In our frame of reference, the jet velocity is presented by $\vec{v}_0 = (0, 0, v_0)$, and the homogeneous ambient magnetic field is $\vec{B}_\mathrm{e} = (0, 0, B_\mathrm{e})$.

Dispersion relations of MHD waves propagating on cylindrical untwisted and twisted plasma jets are generally well-known and here we will only give their final form---their derivation the reader can see in \citealp{zhel12a,zhel,zaq14} and references therein.  We recall that all perturbed quantities associated with the waves are proportional to $\exp[-\mathrm{i}(\omega t -m \phi - k_z z)]$, where $\omega$ is the wave angular frequency, $m$ the azimuthal wave mode number, and $k_z$ the axial wavenumber (as usual, we assume that waves propagate in $z$ direction).  The dispersion relation of transverse MHD mode propagating on an untwisted moving magnetic flux tube of compressible plasma surrounded by also compressible fully ionized plasma has the form \citep{terr,nak,zhel12a}
\begin{eqnarray}
\label{eq:dispeq}
	\frac{\rho_\mathrm{e}}{\rho_\mathrm{i}}\left( \omega^2 - k_z^2 v_\mathrm{Ae}^2
        \right) m_{0\mathrm{i}}\frac{I_m^{\prime}(m_{0\mathrm{i}}a)}{I_m(m_{0\mathrm{i}}a)} \nonumber \\
        \nonumber \\
        {}- \left[ \left( \omega - \vec{k} \cdot
        \vec{v}_0 \right)^2 - k_z^2 v_\mathrm{Ai}^2 \right] m_{0\mathrm{e}}\frac{K_m^{\prime}(m_{0\mathrm{e}}a)}{K_m(m_{0\mathrm{e}}a)} = 0,
\end{eqnarray}
where $v_\mathrm{Ai,e} = B_\mathrm{i,e}/\sqrt{\mu \rho_\mathrm{i,e}}$ are the Alfv\'en speeds in both media, $I_m$ and $K_m$ are modified Bessel functions of the first and second kind, respectively, and the prime implies differentiation with respect to function argument.  As seen, inside the moving flux tube the wave frequency is Doppler-shifted.  The propagation waves are in general surface modes and their attenuation coefficients inside the jet (label `i') and in its environment (label `e') are given by the expression
\begin{equation}
\label{eq:m0sqr}
    m_0^2 = -\frac{\left( \Omega^2 - k_z^2 c_\mathrm{s}^2 \right)\left( \Omega^2 - k_z^2 v_\mathrm{A}^2 \right)}{\left(
    c_\mathrm{s}^2 + v_\mathrm{A}^2 \right)\left( \Omega^2 - k_z v_\mathrm{T}^2 \right)},
\end{equation}
where $\Omega = \omega - \vec{k}\cdot \vec{v}_0$ is the Doppler-shifted mode frequency, $c_\mathrm{si,e} = (\gamma p_\mathrm{i,e}/\rho_\mathrm{i,e})^{1/2}$ with $\gamma = 5/3$ are the sound speeds, and $v_\mathrm{T}$ is so called tube speed defined by the expression \citep{edw}
\[
    c_\mathrm{T} = \frac{c_\mathrm{s}v_\mathrm{A}}{\sqrt{c_\mathrm{s}^2 + v_\mathrm{A}^2}}.
\]

We note that in the limit of incompressible plasmas ($c_\mathrm{si,e} \to \infty$) the wave attenuation coefficients in both media become equal to the axial wavenumber $k_z$ and the wave dispersion relation (\ref{eq:dispeq}) takes the simple form of a quadratic equation, that provides solutions for the real and imaginary part of of the wave phase velocity in closed forms \citep{zhel12b,zhel13}:
\begin{equation}
\label{eq:vph}
    v_\mathrm{ph} \equiv \frac{\omega}{k_z} = \frac{-M_\mathrm{A}B \pm \sqrt{D}}{\eta A - B}v_\mathrm{Ai},
\end{equation}
where $M_\mathrm{A} = v_0/v_\mathrm{Ai}$ is the Alfv\'en Mach number,
\[
    A = I_m^{\prime}(k_z a)/I_m(k_z a), \qquad B = K_m^{\prime}(k_z a)/K_m(k_z a),
\]
and the discriminant $D$ is
\[
    D = M_\mathbf{A}^2 B^2 - (\eta A - B)[(1 - M_\mathbf{A}^2)B - Ab^2].
\]
Obviously, if $D \geqslant 0$, then
\[
    \mathrm{Re}(v_\mathrm{ph}) = \frac{-M_\mathrm{A}B \pm \sqrt{D}}{\eta A - B}v_\mathrm{Ai}, \qquad \mathrm{Im}(v_\mathrm{ph}) = 0,
\]
else
\[
    \mathrm{Re}(v_\mathrm{ph}) = \frac{-M_\mathrm{A}B}{\eta A - B}v_\mathrm{Ai}, \qquad \mathrm{Im}(v_\mathrm{ph}) = \frac{\sqrt{-D}}{\eta A - B}v_\mathrm{Ai}.
\]
We point out that our choice of the sign of $\sqrt{-D}$ in the expression for $\mathrm{Im}(v_\mathrm{ph})$ is plus although, in principle, it might also be minus---in that case, owing to arising instability the wave's energy is transferred to the jet.

For the kink ($m = 1$) mode one can define, for a static magnetic flux tube, the kink speed \citep{edw}
\begin{equation}
\label{eq:kinkspeed}
	c_\mathrm{k} = \left( \frac{\rho_\mathrm{i} v_\mathrm{Ai}^2 + \rho_\mathrm{e}
        v_\mathrm{Ae}^2}{\rho_\mathrm{i} + \rho_\mathrm{e}} \right)^{1/2} = \left(
        \frac{1 + B_\mathrm{e}^2/B_\mathrm{i}^2}{1 +
        \rho_\mathrm{e}/\rho_\mathrm{i}} \right)^{1/2}v_\mathrm{Ai},
\end{equation}
which is independent of sound speeds and characterizes the propagation of transverse perturbations.  As we will demonstrate shortly, the kink mode can become unstable against the KH instability.

As one can expect, the dispersion relation of MHD modes propagating on a moving twisted flux tube is more complicated and has the form \citep{zhel,zaq14}
\begin{eqnarray}
\label{eq:twdispeq}
	\frac{\left( \Omega^2 -
    \omega_\mathrm{Ai}^2 \right)F_m(\kappa_\mathrm{i}a) - 2mA \omega_\mathrm{Ai}/\sqrt{\mu \rho_\mathrm{i}}}
    {\left( \Omega^2 -
    \omega_\mathrm{Ai}^2 \right)^2 - 4A^2\omega_\mathrm{Ai}^2/\mu \rho_\mathrm{i} } \nonumber \\
    \nonumber \\
    {} = \frac{P_m(k_z a)}
    {{\displaystyle \frac{\rho_\mathrm{e}}{\rho_\mathrm{i}}} \left( \omega^2 - \omega_\mathrm{Ae}^2
    \right) + A^2  P_m(k_z a)/\mu \rho_\mathrm{i}},
\end{eqnarray}
where
\begin{equation}
\label{eq:alfvenfrq}
    \omega_\mathrm{Ai} = \frac{mA + k_z B_{\mathrm{i}z}}{\sqrt{\mu \rho_\mathrm{i}}} \quad \mbox{and} \quad \omega_\mathrm{Ae} = \frac{k_z B_{\mathrm{e}z}}{\sqrt{\mu \rho_\mathrm{e}}} = k_z v_{\rm Ae}
\end{equation}
are the local Alfv\'en frequencies inside the moving flux tube and its environment, respectively,
\[
    F_m(\kappa_\mathrm{i}a) = \frac{\kappa_\mathrm{i}a I_m^{\prime}(\kappa_\mathrm{i}a)}{I_m(\kappa_\mathrm{i}a)},
\]
and
\[
    P_m(k_z a) = \frac{k_z a K_m^{\prime}(k_z a)}{K_m(k_z a)}.
\]
The wave attenuation coefficients in Bessel functions' arguments in this case have the forms
\[
    \kappa_\mathrm{i} = k_z \left[  1 - \frac{4 A^2 \omega_\mathrm{Ai}^2}
        {\mu \rho_\mathrm{i} \left( \Omega^2 -
        \omega_\mathrm{Ai}^2\right)^2} \right]^{1/2} \quad \mbox{and} \quad \kappa_\mathrm{e} = k_z.
\]

We note that in the case of a twisted magnetic flux tube with non-magnetized environment dispersion equation (\ref{eq:twdispeq}) recovers Eq.~(13) in \citealp{zaq10}.

\section{Numerical solutions to dispersion relations}
\label{sec:numerics}
Both wave dispersion relations (\ref{eq:dispeq}) and (\ref{eq:twdispeq}) are transcendent equations in which the wave frequency, $\omega$, is a complex quantity: $\mathrm{Re}(\omega) + \mathrm{Im}(\omega)$, where the real and imaginary parts correspond to the frequency and the growth rate of unstable MHD modes, respectively, while the axial wavenumber, $k_z$, is a real variable.  We will solve first Eq.~(\ref{eq:dispeq}) governing the wave propagation in an untwisted moving flux tube.  The occurrence of the expected KH instability is determined primarily by the jet velocity and in searching for a critical or threshold value of it, we will gradually change velocity magnitude from zero to that critical value (and beyond). Thus, we have to solve the dispersion relation in complex variables, obtaining the real and imaginary parts of the wave frequency, or as is commonly accepted, of the wave phase velocity $v_\mathrm{ph} = \omega/k_z$, as functions of $k_z$ at various values of the velocity shear between the EUV jet and its environment, $\mathbf{v}_0$.  We emphasize that for our piecewise uniform plasma density profile the singularities of the Alfv\'enic continuous spectrum that would exist if the density were continuously varying are all concentrated in the point
$r = a$ \citep{goos}.  Note that this is true for all modes with $m \neq 0$.  Moreover, since we are interested in unstable MHD waves, they must be discrete modes (see Chap.~8 and Fig.~8.4 in \citealp{jard}).

It is well known that the normal modes propagating along a bounded homogeneous magnetized cylindrical plasma column can be pure surface waves, pseudosurface (body) waves, or leaky waves \citep{call}.  The type of the wave crucially depends on the ordering of the basic speeds in both media (the column and its surrounding plasma), more specifically sound and Alfv\'en speeds as well as corresponding tube speeds.  Borrowing for a while Cally's notation ($a$ stands for internal Alfv\'en speed, $a_\mathrm{e}$ for external Alfv\'en one, $c$ for internal sound speed, and $c_\mathrm{e}$ for the external one) the speeds' ordering in our case is
\[
    a < a_\mathrm{e} < c < c_\mathrm{e}.
\]
After normalizing all velocities with respect to the external sound speed, $c_\mathrm{e}$, we obtain the dimensionless Alfv\'en, sound, and tube speeds as follows:
\[
    A = 0.2865, \quad C = 0.771, \quad A_\mathrm{e} = 0.3735,
\]
\[
    C_\mathrm{T} = 0.2886, \quad \mbox{end} \quad C_\mathrm{Te} = 0.35.
\]
In calculating $C_\mathrm{T}$ and $C_\mathrm{Te}$ we have used $c_\mathrm{Ti} = 44.59$~km\,s$^{-1}$ and $c_\mathrm{Te} = 58.08$~km\,s$^{-1}$, respectively.  Thus we have the following relations:
\[
    A < C \quad \mbox{and} \quad A < C_\mathrm{Te}, \quad \mbox{as well as} \quad A < C_\mathrm{Te} < C.
\]
Now looking at Table I in \cite{call}, we conclude that the kink ($m = 1$) mode must be a non-leaky pure surface mode of type S$_{+}^{-}$ that implies an externally slow internally fast surface wave.  By `fast' and 'slow' here, it is meant that the normalized wave phase velocity $V$ is either greater than or less than the lesser of the sound and Alfv\'en speeds.  For an S$_{+}^{-}$ type wave $V$ should lies between $A$ and $C$: $A \leqslant V \leqslant C$, and also $V < C_\mathrm{Te}$.  We note that the first inequalities' chain in Table I is wrong---the correct one is listed above.  It is curious to see whether the numerical solving Eq.~(\ref{eq:dispeq}) will confirm these predictions.

Before starting the numerical task, we have to normalize all variables and to specify the input parameters.  Our normalization differs from that of \cite{call}---we find more convenient the wave phase velocity, $v_\mathrm{ph}$, and the other speeds to be normalized with respect to the Alfv\'en speed inside the jet, $v_\mathrm{Ai}$---Cally's normalization is inapplicable for the case when the surrounding plasma is considered as an incompressible medium.  The wavelength, $\lambda = 2\pi/k_z$, is normalized to the tube radius, $a$, that is equivalent to introducing a dimensionless wavenumber $k_z a$.  For normalizing the Alfv\'en speed in the ambient coronal plasma, $v_\mathrm{Ae}$, we need the density contrast, $\eta$, and the ratio of the magnetic fields $b = B_\mathrm{e}/B_\mathrm{i}$, to get $v_\mathrm{Ae}/v_\mathrm{Ai} = b/\sqrt{\eta}$.  The normalization of sound speeds in both media requires the specification of the reduced plasma betas, $\tilde{\beta}_\mathrm{i,e} = c_\mathrm{si,e}/v_\mathrm{Ai,e}$.  In the dimensionless analysis the flow speed, $v_0$, will be presented by the Alfv\'en Mach number $M_\mathrm{A} = v_0/v_\mathrm{Ai}$.
The input parameters for solving Eq.~(\ref{eq:dispeq}) are as follows: $\eta = 0.589$, $b = 1.014$, $\tilde{\beta}_\mathrm{i} = 7.3$, and $\tilde{\beta}_\mathrm{e} = 7.07$; during the calculations we will vary the Alfv\'en Mach number, $M_\mathrm{A}$, from zero (static plasma) to values at which we will obtain unstable solutions.  Among the various MHD modes propagating on a cylindrical static/moving magnetic flux tube of compressible plasma surrounded by also compressible medium the most interesting is the kink ($m = 1$) mode because, as previous studies \citep{and,vash,zhel12a,zhel,zaq14} show, namely the kink mode becomes
unstable when the jet speed exceeds a critical value.  At a static magnetic flux tube ($M_\mathrm{A} = 0$), the kink waves propagates with the kink speed, $c_\mathrm{k}$ (see Eq.~(\ref{eq:kinkspeed})), which in our case is equal to $53.6$~km\,s$^{-1}$, or in dimensionless units, $1.1269$, i.e., the wave is slightly super-Alfv\'enic one and $v_\mathrm{ph}$ lies between the internal and external Alfv\'en speeds.  This is not surprising because the kink speed (\ref{eq:kinkspeed}) is not sensitive to the sound speed---that is why it is more natural $C$ in the inequalities' chain in Cally's \citep{call} Table I to be replaced by $A_\mathrm{e}$.  Our numerical calculation confirm that $v_\mathrm{ph} < c_\mathrm{Te}$.  In Cally's dimensionless variables our normalized kink speed of $1.1269$ transforms into $0.3228$ and we have $0.2865 < 0.3228 < 0.3735$ along with $0.3228 < 0.35$, that is, in agreement with aforementioned predictions.  One can also conclude that the way of normalization is not too decisive---even with our normalization we get the correct inequalities' chain.  Calculated during the solving transcendental Eq.~(\ref{eq:dispeq}) wave attenuation coefficients $m_{0\mathrm{i,e}}$ were (for $M_\mathrm{A} = 0$) real quantities confirming that the principal kink mode is a pure surface wave.
\begin{figure}[!ht]
  \centering
\subfigure{\includegraphics[width = 3.3in]{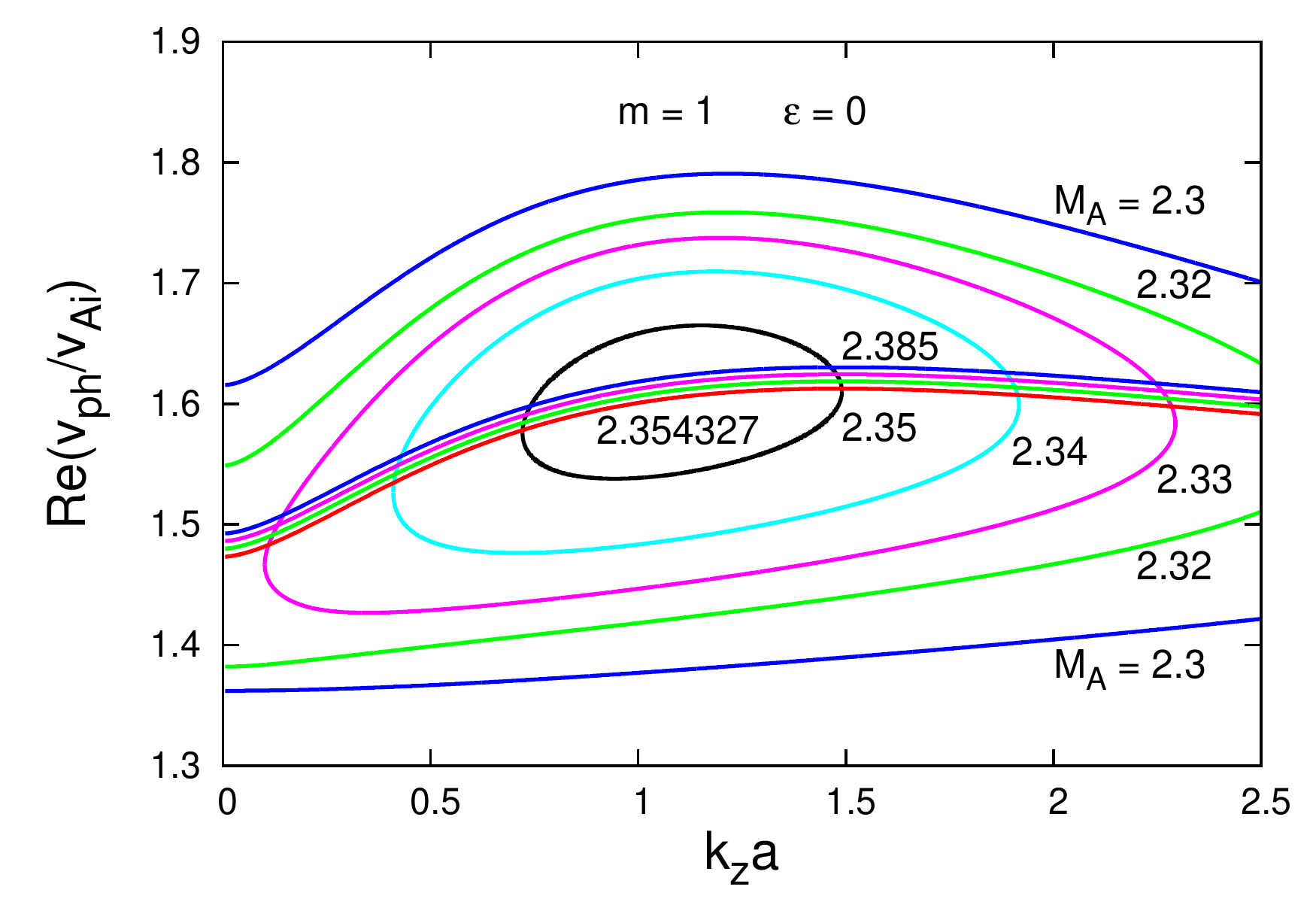}} \\
\subfigure{\includegraphics[width = 3.3in]{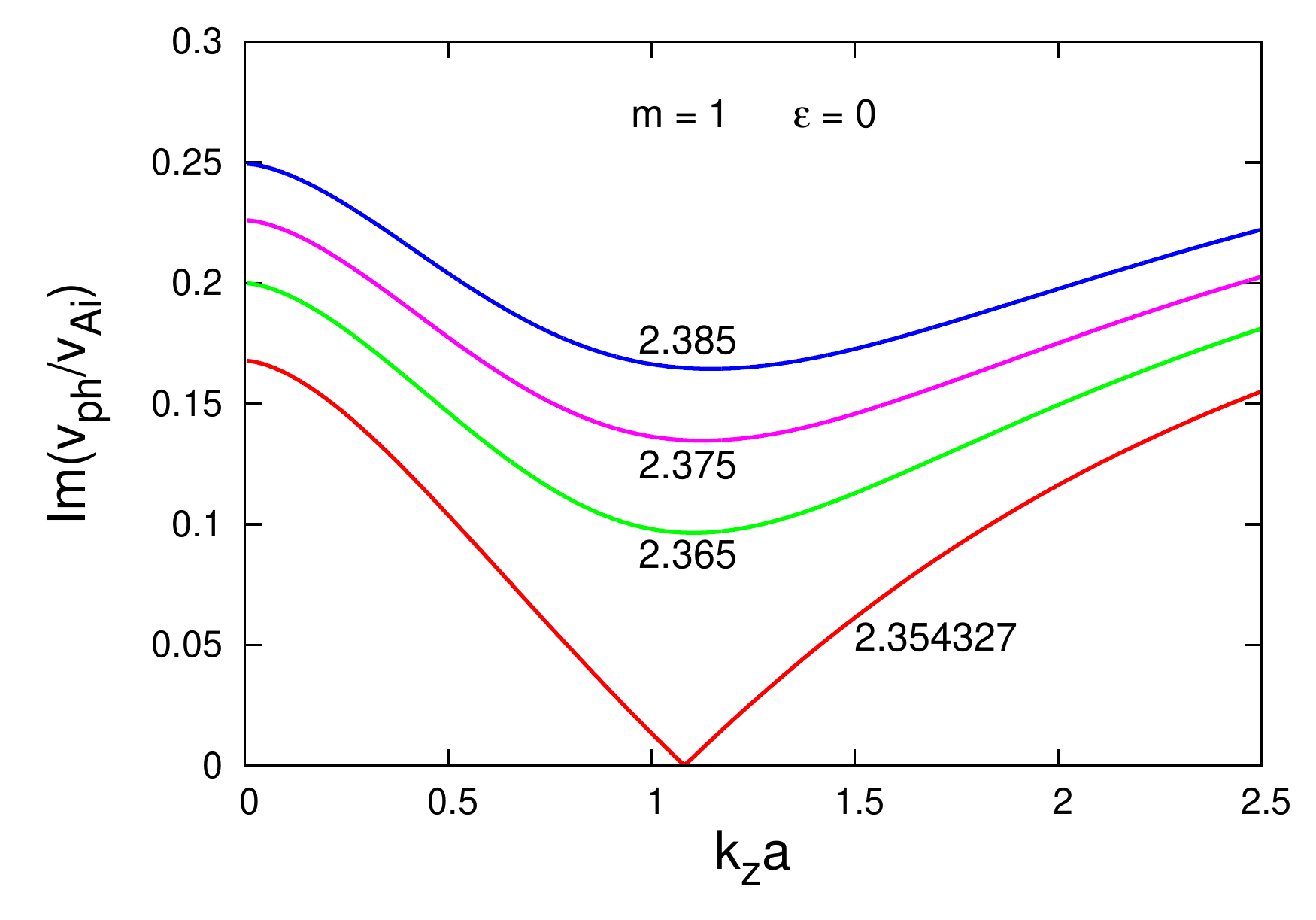}}
  \caption{(\emph{Top panel}) Dispersion curves of stable and unstable kink ($m = 1$) MHD mode propagating on a moving untwisted flux tube of compressible plasma at $\eta = 0.598$ and $b = 0.36$.  Unstable are the waves with dispersion curves located in the middle of the plot for four values of the Alfv\'en Mach number $M_\mathrm{A} = 2.354327$, $2.365$, $2.375$, and $2.385$.  All other curves correspond to stable kink waves. (\emph{Bottom panel})  The normalized growth rates of unstable waves for the same values of $M_\mathrm{A}$.  Red curves in both plots correspond to the onset of KH instability.}
  \label{fig:fig2}
\end{figure}
\begin{figure}[!ht]
  \centering
\subfigure{\includegraphics[width = 3.3in]{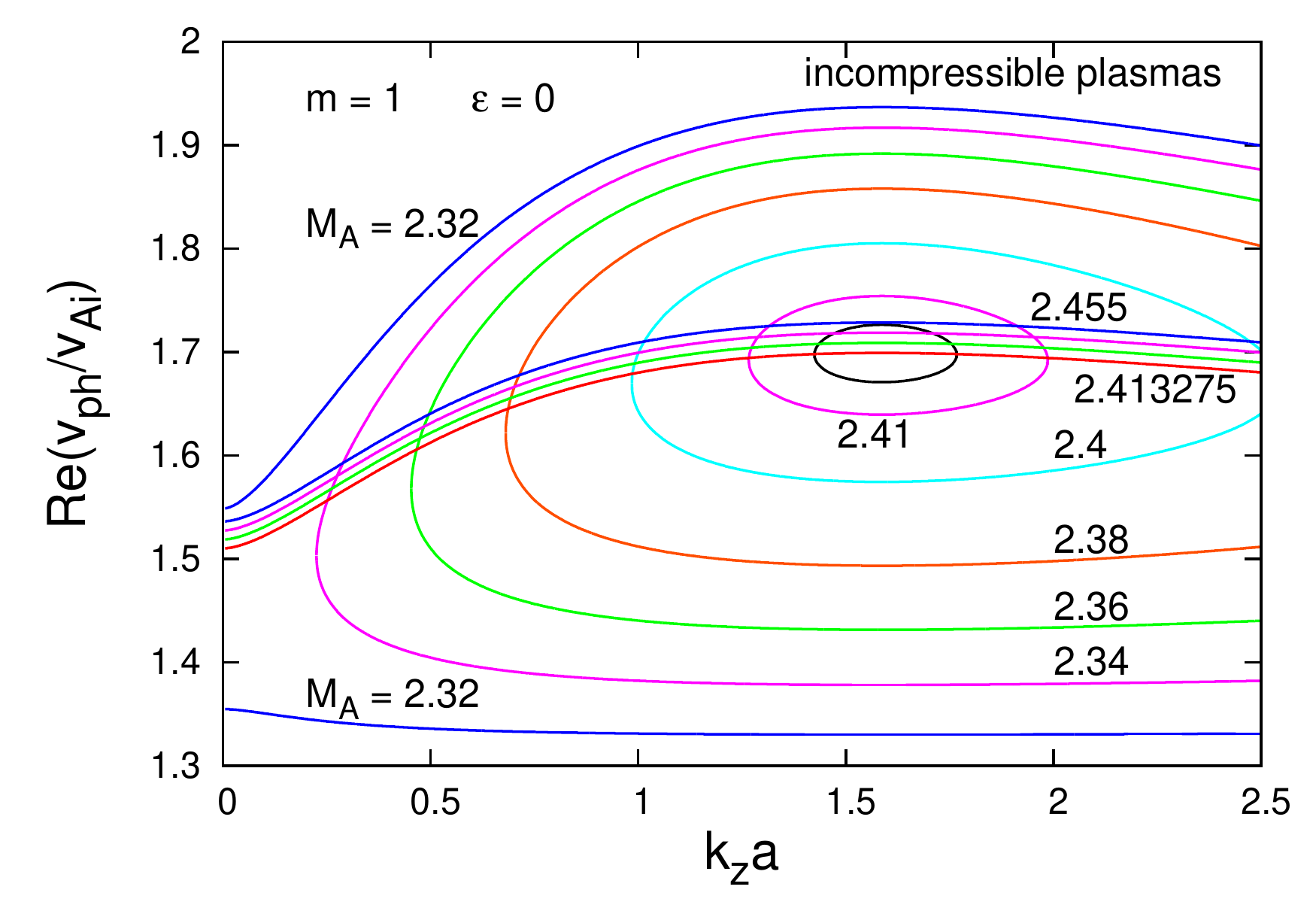}} \\
\subfigure{\includegraphics[width = 3.3in]{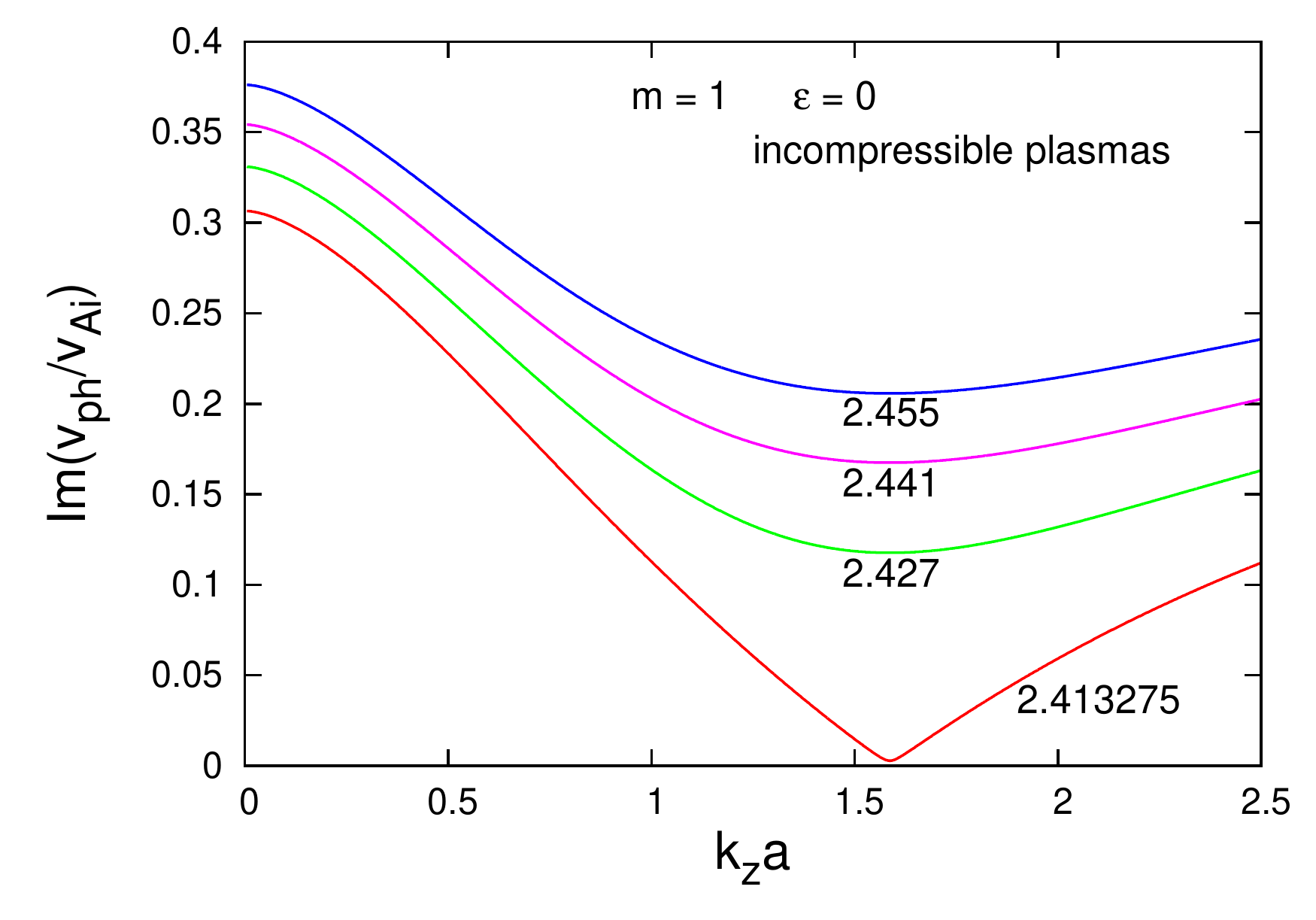}}
  \caption{(\emph{Top panel}) The same as in Fig.~\ref{fig:fig2}, however, for incompressible plasma in both media.  Unstable are the waves for four values of the Alfv\'en Mach number $M_\mathrm{A} = 2.413275$, $2.427$, $2.441$, and $2.455$.  The closed black curve corresponds to $M_\mathrm{A} = 2.4125$.  All other curves correspond to stable kink waves. (\emph{Bottom panel})  The normalized growth rates of unstable waves for the same values of $M_\mathrm{A}$.  Red curves in both plots correspond to the onset of KH instability.}
  \label{fig:fig3}
\end{figure}

The flow shifts upwards the kink-speed dispersion curve as well as splits it into two separate curves, which for small Alfv\'en Mach numbers travel with velocities $M_\mathrm{A} \mp c_\mathrm{k}/v_\mathrm{Ai}$ \citep{zhel12a}.  (A similar duplication happens for the tube-speed dispersion curves, too.)  At higher $M_\mathrm{A}$, however, the behavior of each curve of the pair $M_\mathrm{A} \mp c_\mathrm{k}/v_\mathrm{Ai}$ is completely different.  As seen from the top panel of Fig.~\ref{fig:fig2}, for $M_\mathrm{A} \geqslant 2.33$ both kink curves merge forming a closed dispersion curve.  With increasing $M_\mathrm{A}$ the closed dispersion curves become smaller and narrower and the kink mode becomes unstable at $M_\mathrm{A}^\mathrm{cr} = 2.354327$.  The growth rates of unstable kink modes are plotted in the bottom panel of Fig.~\ref{fig:fig2}.  The red curves in all diagrams denote the marginal dispersion/growth rate curves: for $M_\mathrm{A} < M_\mathrm{A}^\mathrm{cr}$ the kink waves are stable, otherwise they are unstable and the instability is of the KH type.  With $M_\mathrm{A}^\mathrm{cr} = 2.354327$ the kink mode will be unstable if the velocity of the moving flux tube exceeds $112$~km\,s$^{-1}$---a speed, which is below the observationally evaluated jet speed of $150$~km\,s$^{-1}$.
It is important to underline that all the stable kink modes in our system are pure surface waves but the unstable ones are not---the latter become leaky waves (their external attenuation coefficients, $m_{0\mathrm{e}}$s, are complex quantities with positive imaginary parts).  This implies that wave energy is radiated outward in the surrounding medium.  Thus, the KH instability plays a dual role: once in its nonlinear stage the instability can trigger wave turbulence and simultaneously the propagating KH-wave is radiating its energy outside.

The circumstance that the reduced plasma betas in both media are ${\sim}7$, tempts us to consider the moving get and its surrounding magnetized plasma as incompressible media.  Then, as it was discussed in Sect.~\ref{sec:basic}, the wave dispersion equation (\ref{eq:dispeq}) becomes quadratic one giving us the wave phase velocity and its growth rate when the kink mode is unstable in closed forms.  Calculated dispersion curves and growth rates are plotted in Fig.~\ref{fig:fig3}.  One is immediately seen that the picture is very similar to that in Fig.~\ref{fig:fig2}.  Note that both the dispersion curve of stable kink waves and the growth rate of the marginal unstable wave are shifted to the right.  Moreover, the critical Alfv\'en Mach number now is a little bit higher, $M_\mathrm{A}^\mathrm{cr} = 2.413275$ that yields a critical flow speed of $114.8$~~km\,s$^{-1}$.  Unlike the unstable kink mode in compressible plasma, that in incompressible limit persists as a non-leaky surface wave---the two attenuation coefficients are real quantities.
\begin{figure}[!ht]
  \centering
\subfigure{\includegraphics[width = 3.3in]{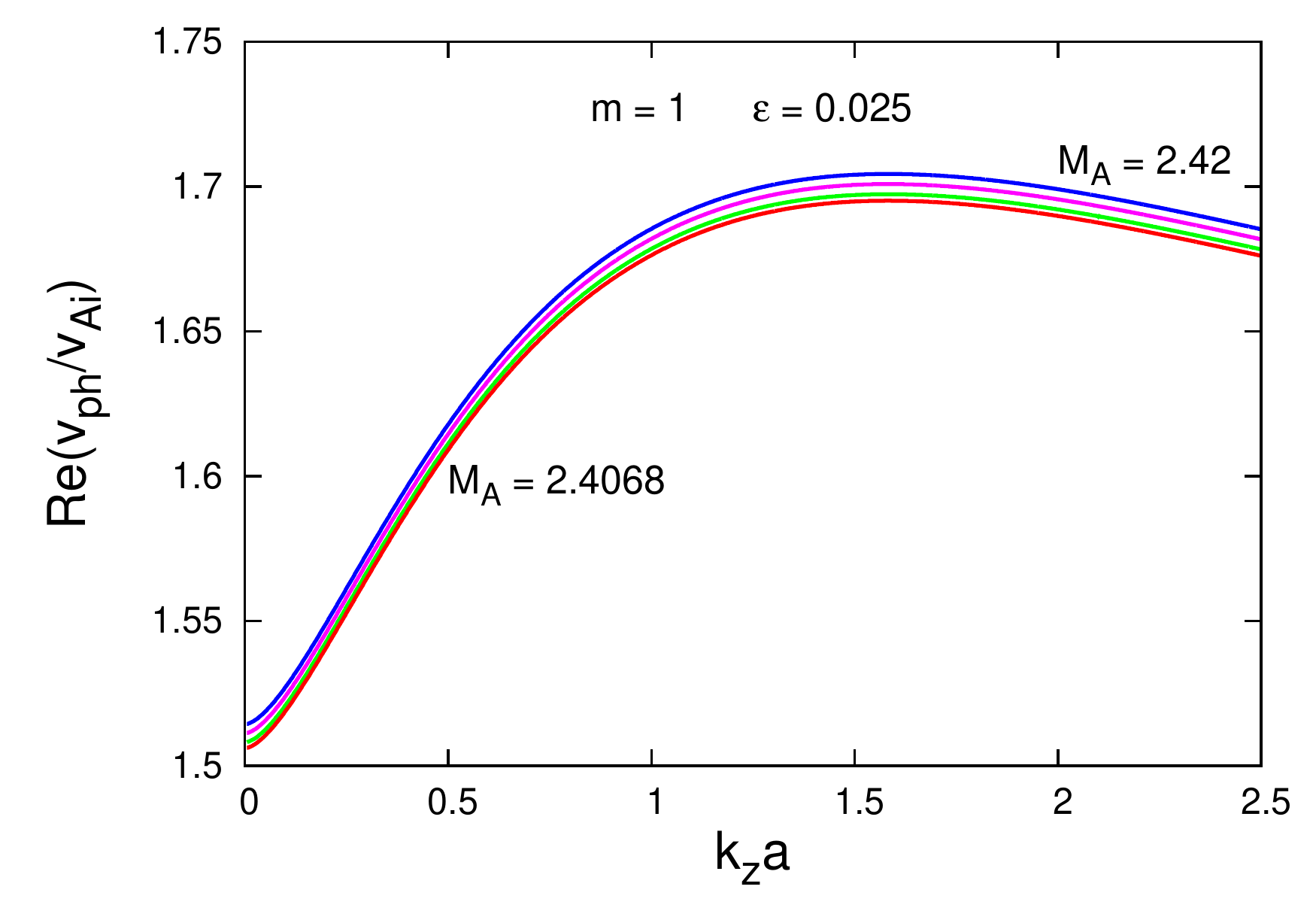}} \\
\subfigure{\includegraphics[width = 3.3in]{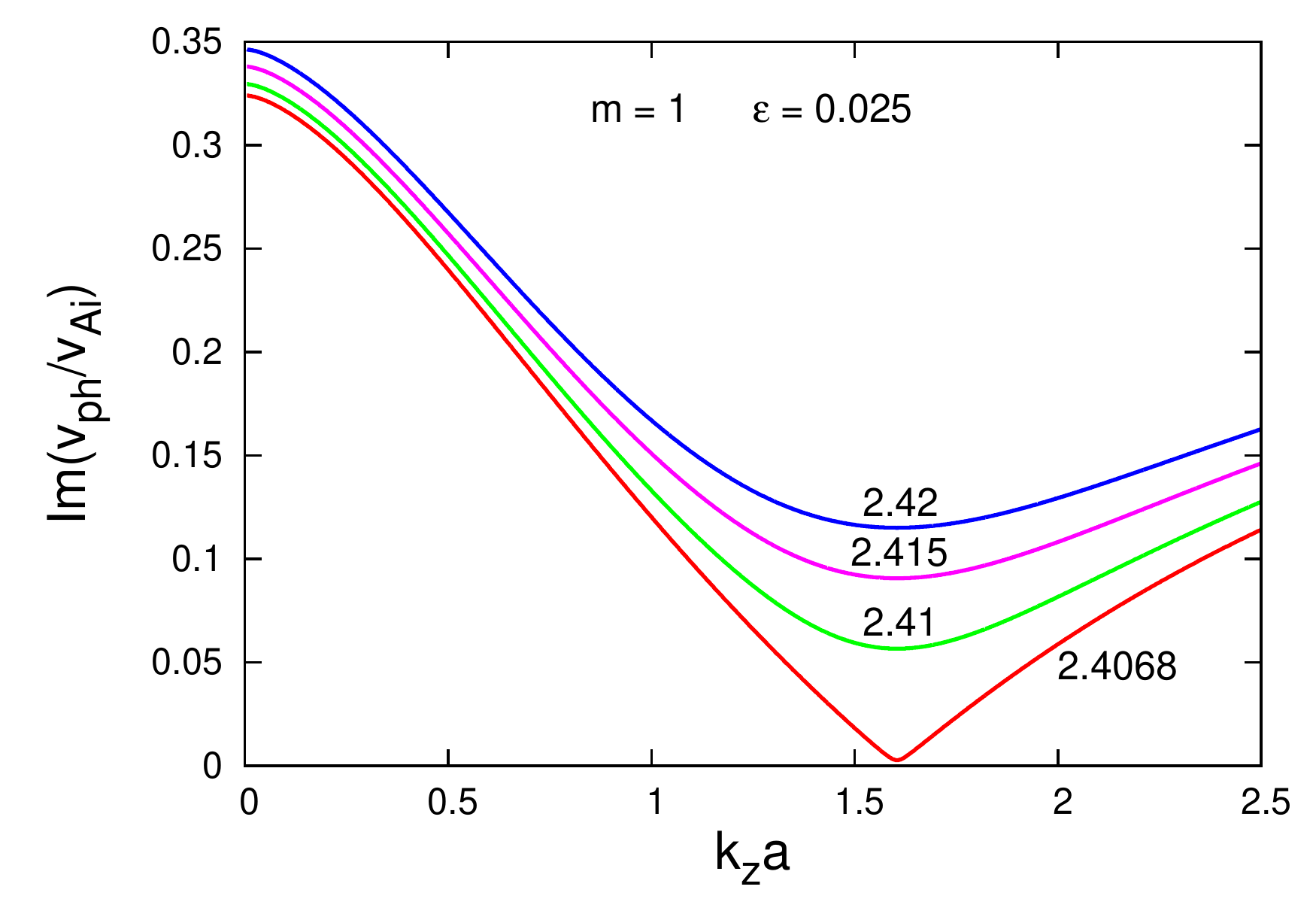}}
  \caption{(\emph{Top panel}) Dispersion curves of unstable kink ($m = 1$) MHD mode propagating on a moving twisted flux tube of incompressible plasma at $\varepsilon = 0.025$, and for four values of the Alfv\'en Mach number $M_\mathrm{A} = 2.4068$, $2.41$, $2.415$, and $2.42$. (\emph{Bottom panel})  The normalized growth rates for the same values of $M_\mathrm{A}$.  Red curves in both plots correspond to the onset of KH instability.}
  \label{fig:fig4}
\end{figure}

When we model the EUV jet as a moving twisted magnetic flux tube two types of instabilities can develop in the jet, notably kink instability due to the twist of the magnetic field and KH instability owing to the tangential discontinuity of plasma flow at the tube boundary.  A normal mode analysis \citep{dung} and an energy consideration method (calculating the change in magnetic energy per unit length of the cylinder---see \citealp{lund}) yield similar thresholds of the kink instability in twisted magnetic flux tubes in the form $B_{\mathrm{i}\phi}(a) > 2 B_{\mathrm{i}z}$, or in our notation, as $\varepsilon > 2$.  In the following, we will consider only weakly twisted tubes, $\varepsilon \ll 1$, and, therefore, only the KH instability may occur in the exploring jet configuration.  In solving Eq.~(\ref{eq:twdispeq}) we use the same method as in the case of untwisted tube.  The new quantities for normalization that appear here are the local Alfv\'en frequencies $\omega_\mathrm{Ai,e}$.  The natural way of normalizing  $\omega_\mathrm{Ai}$ is to multiply it by the tube radius, $a$, and after that divide by the Alfv\'en speed $v_\mathrm{Ai}$, that is,
\[
    a\omega_\mathrm{Ai} = \frac{B_{\mathrm{i}z}}{\sqrt{\mu \rho_\mathrm{i}}}\left( m\frac{Aa}{B_{\mathrm{i}z}} + k_z a \right),
\]
or
\[
     \frac{a\omega_\mathrm{Ai}}{v_\mathrm{Ai}} = m\frac{Aa}{B_{\mathrm{i}z}} + k_z a = m\varepsilon + k_z a.
\]
Note that here the Alfv\'en speed is defined via the axial component of the twisted magnetic field, $v_\mathrm{Ai} = B_{\mathrm{i}z}/\sqrt{\mu \rho_\mathrm{i}}$.  Accordingly, the normalized local Alfv\'en frequency $\omega_\mathrm{Ae}$ has the form
\[
    \frac{a\omega_\mathrm{Ae}}{v_\mathrm{Ai}} = k_z a\frac{B_\mathrm{e}/B_{\mathrm{i}z}}{\sqrt{\eta}} = k_z a \frac{b_\mathrm{twist}}{\sqrt{\eta}},
\]
\begin{figure}[!ht]
  \centering
\subfigure{\includegraphics[width = 3.3in]{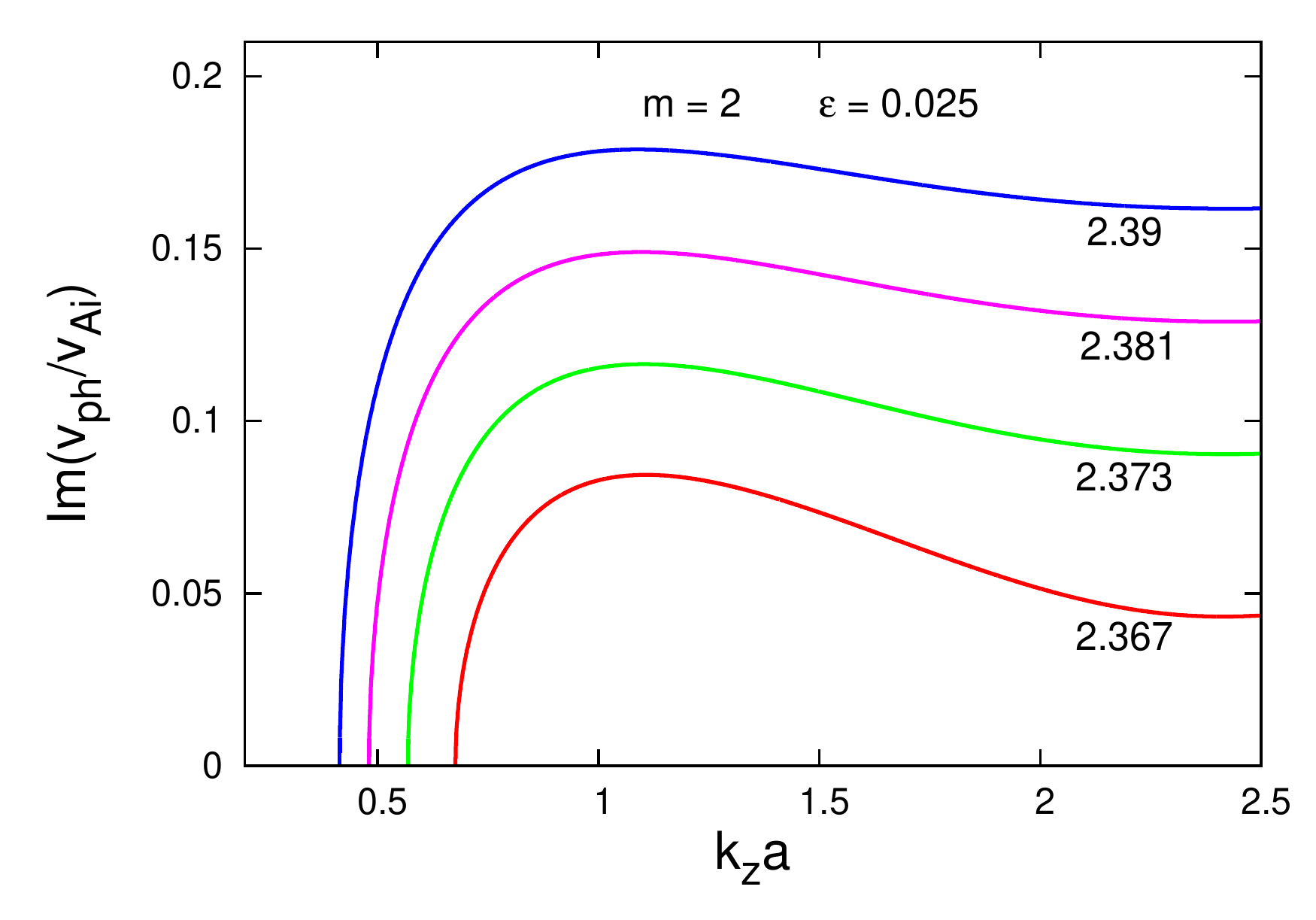}} \\
\subfigure{\includegraphics[width = 3.3in]{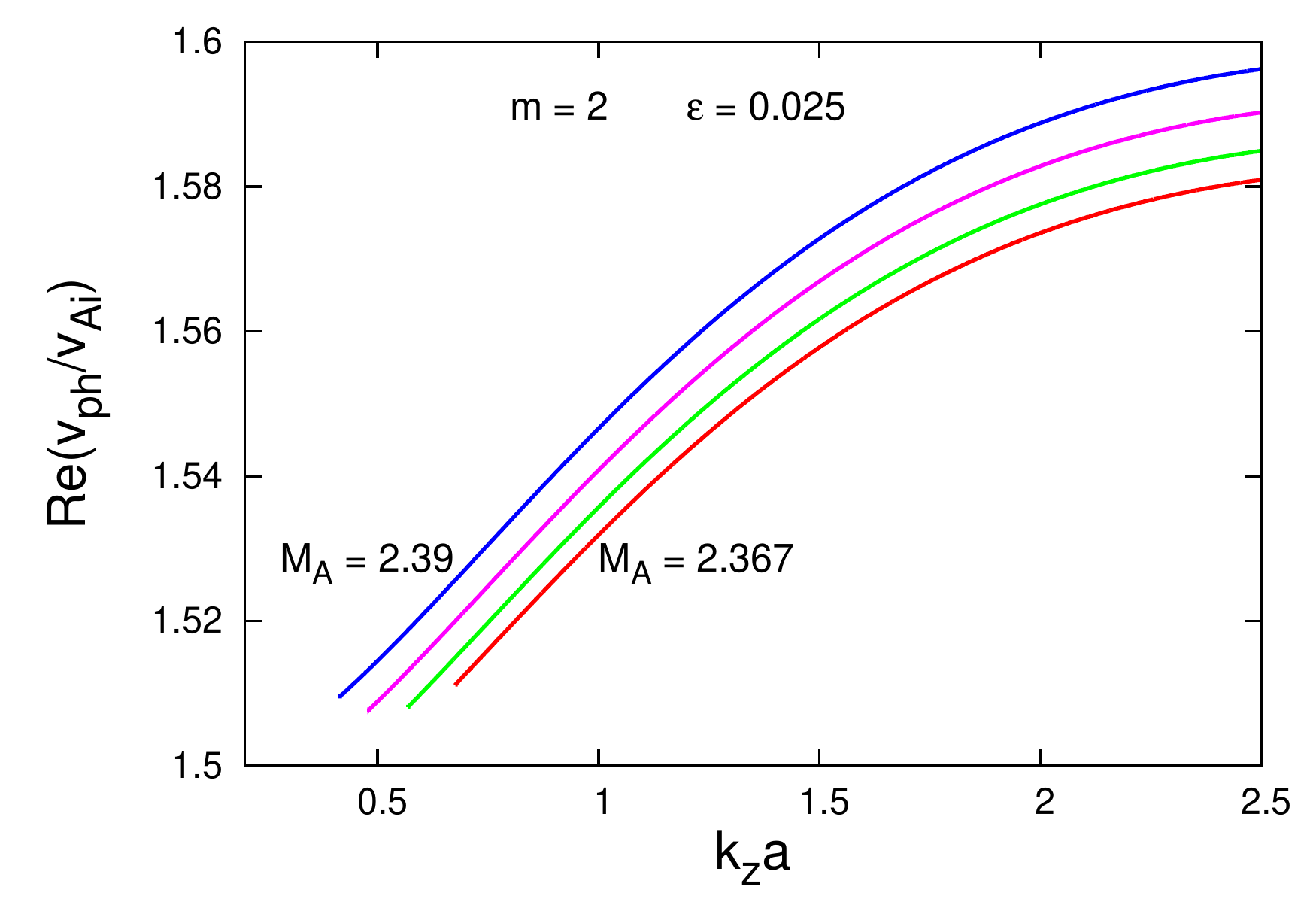}}
  \caption{(\emph{Top panel}) Normalized growth rates of unstable fluting-like ($m = 2$) MHD mode propagating on a moving twisted flux tube of incompressible plasma at $\varepsilon = 0.025$, and for four values of the Alfv\'en Mach number $M_\mathrm{A} = 2.367$, $2.373$, $2.381$, and $2.39$. (\emph{Bottom panel})  Wave dispersion curves for the same values of $M_\mathrm{A}$.  Red curves in both plots correspond to the onset of KH instability.}
  \label{fig:fig5}
\end{figure}
\begin{figure}[!ht]
  \centering
\subfigure{\includegraphics[width = 3.3in]{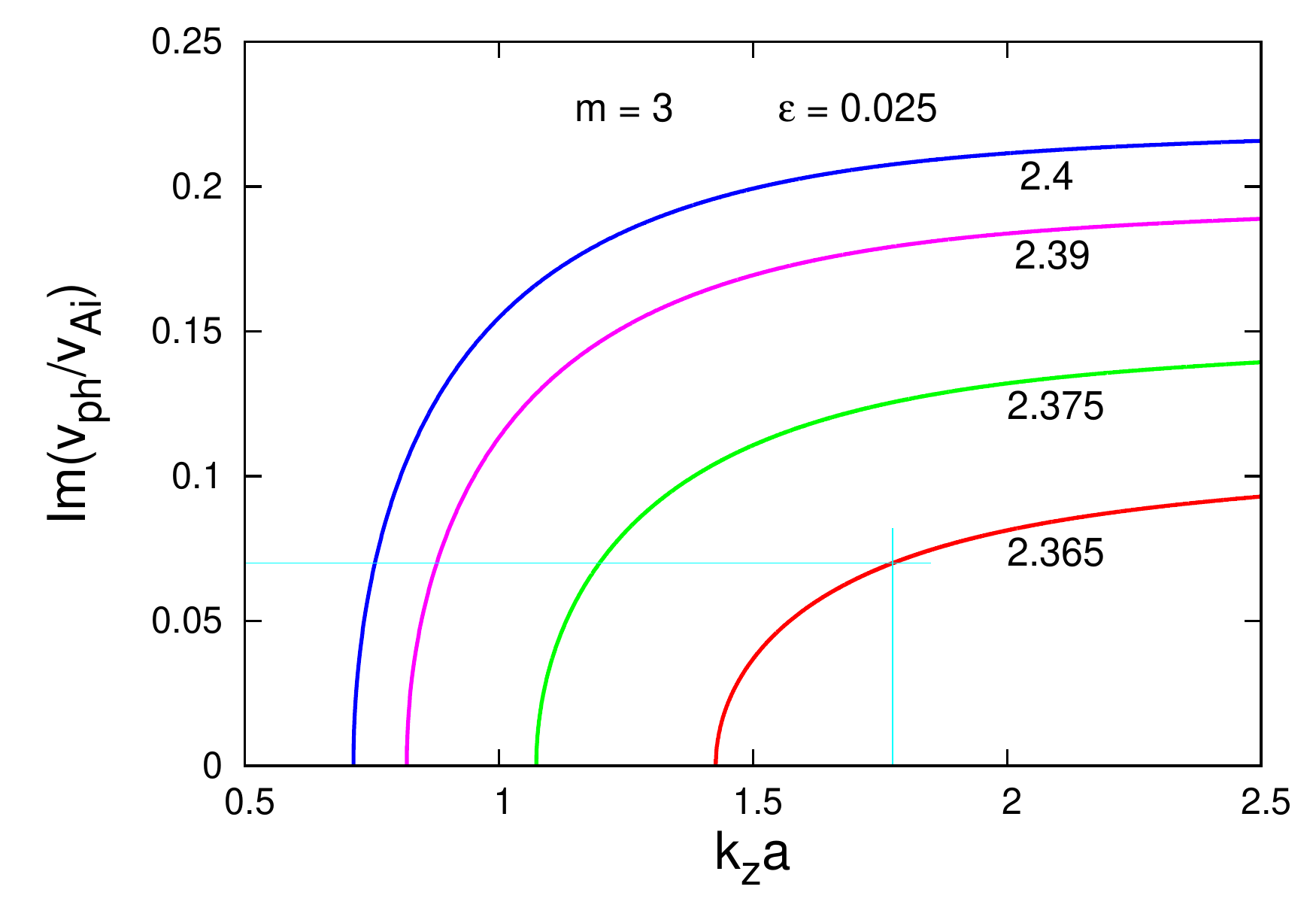}} \\
\subfigure{\includegraphics[width = 3.3in]{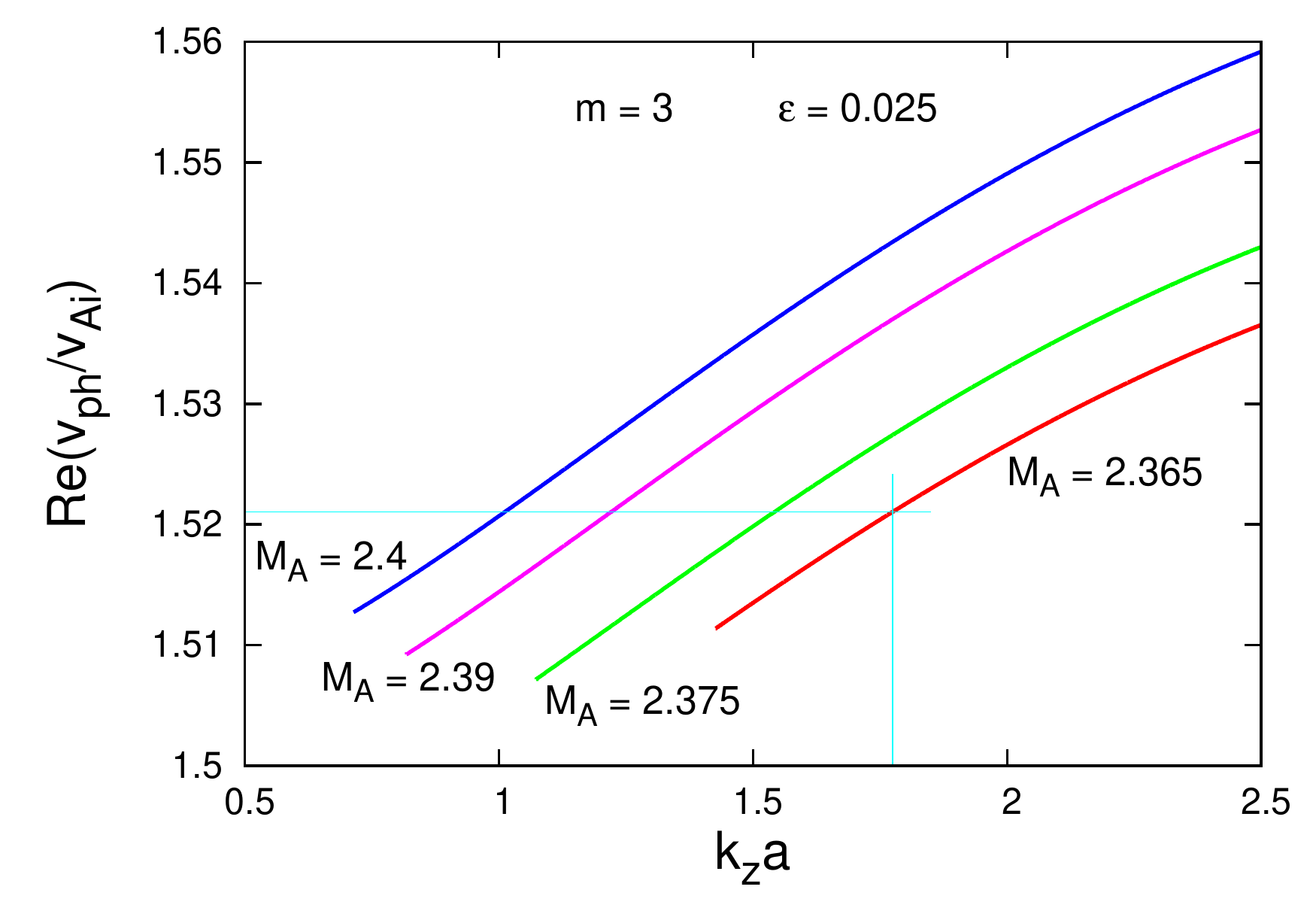}}
  \caption{(\emph{Top panel}) Normalized growth rates of unstable $m = 3$ MHD mode propagating on a moving twisted flux tube of incompressible plasma at $\varepsilon = 0.025$, and for four values of the Alfv\'en Mach number $M_\mathrm{A} = 2.365$, $2.375$, $2.39$, and $2.4$. (\emph{Bottom panel})  Wave dispersion curves for the same values of $M_\mathrm{A}$.  Red curves in both plots correspond to the onset of KH instability.}
  \label{fig:fig6}
\end{figure}
\begin{figure}[!ht]
  \centering
\subfigure{\includegraphics[width = 3.3in]{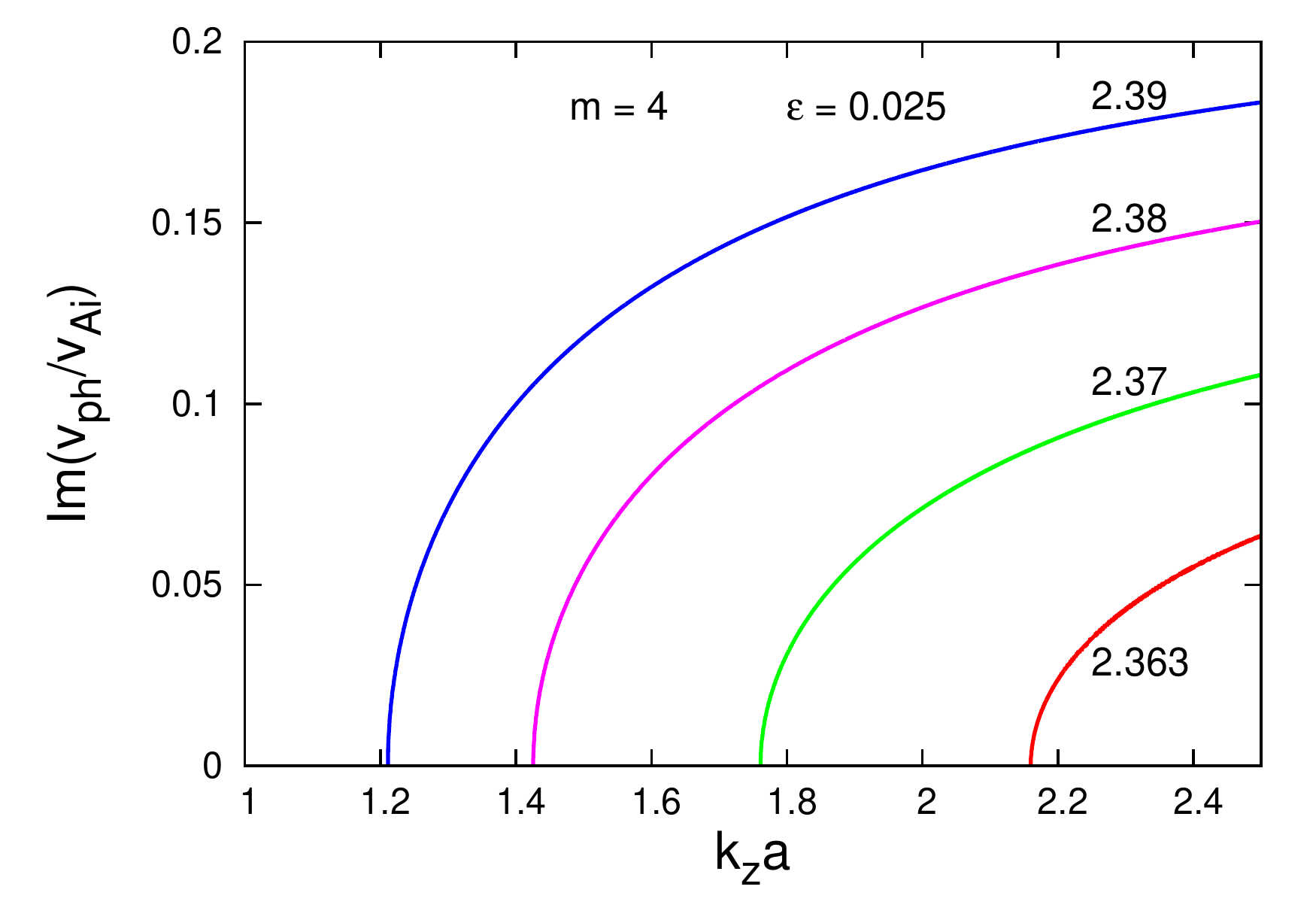}} \\
\subfigure{\includegraphics[width = 3.3in]{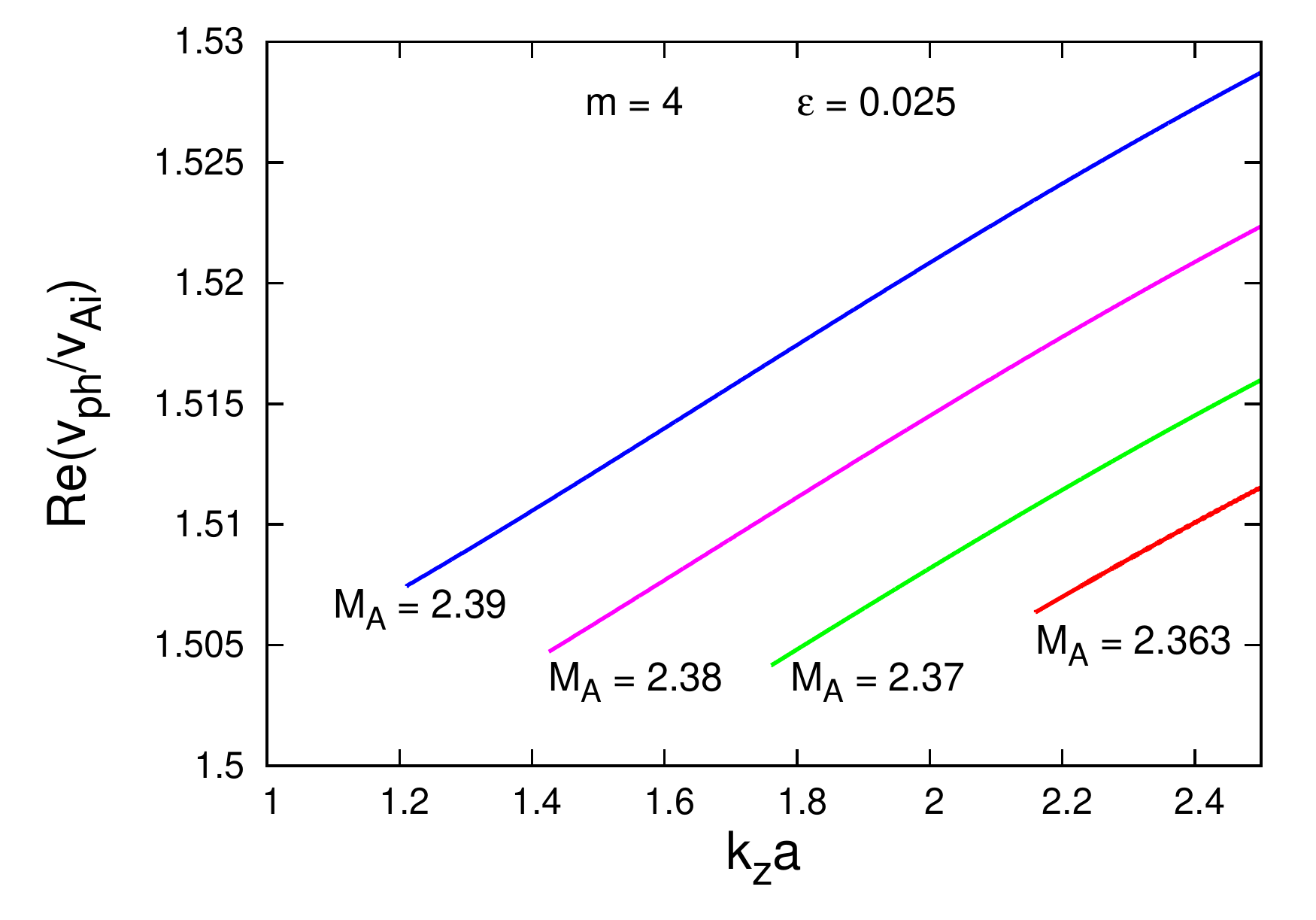}}
  \caption{(\emph{Top panel}) Normalized growth rates of unstable $m = 4$ MHD mode propagating on a moving twisted flux tube of incompressible plasma at $\varepsilon = 0.025$, and for four values of the Alfv\'en Mach number $M_\mathrm{A} = 2.363$, $2.37$, $2.38$, and $2.39$. (\emph{Bottom panel})  Wave dispersion curves for the same values of $M_\mathrm{A}$.  Red curves in both plots correspond to the onset of KH instability.}
  \label{fig:fig7}
\end{figure}
where the parameter $b_\mathrm{twist} = B_\mathrm{e}/B_{\mathrm{i}z} = b\sqrt{1 + \varepsilon^2}$.  For small values of the twist parameter $\varepsilon$, we take $b_\mathrm{twist} \cong b$, and bearing in mind that $b$ has a value close to $1$, one can assume that $b = 1$.  We begin our numerical calculations for the kink ($m = 1$) mode in the moving twisted tube with $\eta = 0.598$, $b_\mathrm{twist} \cong b = 1$, and $\varepsilon = 0.025$.  Dispersion curves and growth rates of unstable waves are plotted in Fig.~\ref{fig:fig4}.  As seen, the curves are very similar to those shown in Figs.~\ref{fig:fig2} and \ref{fig:fig3}, of course, with the observation that the minimum of the marginal growth rate (red) curve of the kink mode in shifted to the right on the $k_z a$-axis as compared with the plot in the bottom panel of Fig.~\ref{fig:fig2} for a similar curve of the $m = 1$ mode traveling on untwisted magnetic flux tube containing compressible plasma.  On the other hand, the minima of the marginal growth rate curves in untwisted and twisted flux tubes in the limits of incompressible plasmas practically coincide (compare the bottom panels in Figs.~\ref{fig:fig3} and \ref{fig:fig4}.  The critical flow velocity for emerging KH instability now is $v_0^\mathrm{cr} = 114.4$~km\,s$^{-1}$, calculated by using the `reduced' Alfv\'en speed $v_\mathrm{Ai}/\sqrt{1 + \varepsilon^2} = 47.522$~km\,s$^{-1}$.  It is worth pointing out that the incompressible plasma approximation in general yields slightly higher threshold Alfv\'en Mach numbers than the model of compressible media.  But the value of $114.4$~km\,s$^{-1}$ is an entirely acceptable and reasonable jet speed.  It is instructive to note that the unstable kink ($m = 1$) mode is a non-leaky surface wave---its external attenuation coefficient, $\kappa_\mathrm{e} = k_z$, is not changed by the instability, but the internal one, $\kappa_\mathrm{i}$, becomes a complex quantity with positive real and imaginary parts.  The same applies to the higher fluting-like $(m = 2$) mode and to the higher $m$ modes with azimuthal mode numbers $3$ and $4$.

It is interesting to see what kind of threshold Alfv\'en Mach numbers we will obtain as consider the excitations of higher MHD mode numbers.  Next three figures show the results of numerical computations for the fluting-like $m = 2$ mode, the $m = 3$, and the $m = 4$ mode.  Here we have a new phenomenon: the KH instabilities start at some critical $k_z a$-numbers along with the corresponding critical Alfv\'en Mach numbers.  As one can see from Figs.~\ref{fig:fig5}--\ref{fig:fig7}, those critical wavenumbers have values of $0.676$ trough $1.427$ to $2.16099$ for the $m = 2$, $m = 3$, and $m = 4$ MHD modes, respectively.  If we assume that our EUV jet has a width $\Delta \ell = 4000$~km, then the corresponding critical wavelengths for a KH instability onset are $\lambda_\mathrm{cr}^{m=2} \cong 18.6$~Mm, $\lambda_\mathrm{cr}^{m=3} \cong 8.8$~Mm, and $\lambda_\mathrm{cr}^{m=3} \cong 5.8$~Mm, respectively.  These critical wavelengths are of the order or less than the height of the EUV gets; for instance, \cite{zhang} observing three jets in the seven extreme-ultraviolet (EUV) filters of the AIA/\emph{SDO}, have reported jets' heights between $12.8$ and $26.8$~Mm; the arithmetic mean diameter of those EUV jets was $3.7$~Mm.  One expects that the actual wavelengths of occurring KH instabilities of the higher $m$ modes should be shorter than the discussed critical wavelengths.   We pay attention to the fact that the threshold Alfv\'en Mach numbers for observing KH instability in the $m = 2$, $m = 3$, and $m =4$ modes are very close, equal to $2.367$, $2.365$, and $2.363$, respectively, and they yield critical flow velocities of $112.5$, $112.4$, and $112.3$~km\,s$^{-1}$.  Thus, we can conclude that all studied MHD modes propagating on the EUV jet in equal measure can become unstable against the KH instability---the required flow velocity shear lies between $112$ and $114.4$~km\,s$^{-1}$.  We note also that the phase velocities of unstable higher $m$ modes are in the range of $72$--$75$~km\,s$^{-1}$, while those of the kink, $m = 1$, mode lie between $70$ and $76.7$~km\,s$^{-1}$ in the case of untwisted jet, and between $71.6$ and $81.6$~km\,s$^{-1}$ for a weakly twisted EUV jet.  An evaluation of the instability growth rate requires the magnitude of the observed/detected wavelength (or equivalently, the $k_z a$-value) and the jet width.  For example, if we assume that the KH instability of the $m = 3$ MHD harmonic rises (see Fig.~\ref{fig:fig6}) at $k_z a = 1.775$ (or with $\lambda_\mathrm{KH} \cong 7.1$~Mm, assuming that $\Delta \ell = 4$~Mm), the corresponding normalized wave phase velocity growth rate is equal to $0.07$ that yields a wave growth rate $\mathrm{Im}(\omega_\mathrm{KH}) \cong 3\times 10^{-3}$~s$^{-1}$.  Such growth rates, of few inverse milliseconds, were evaluated for the KH instability of high harmonic MHD modes (however, with negative signs of their azimuthal wave mode numbers) in cool surges (see, for instance, \citealp{zhel15b}).  The wave phase velocity of the considered unstable $m = 3$ mode (see the bottom panel in Fig.~\ref{fig:fig6}) is equal to ${\approx}72.3$~km\,s$^{-1}$.

\section{Discussion}
\label{sec:discussion}
To numerically study the emergence of KH instability of MHD modes propagating on moving magnetic flux tubes (untwisted and twisted ones) that model our EUV jet, we used the well-known dispersion relation of MHD waves traveling on cylindrical compressible plasmas, Eq.~(\ref{eq:dispeq}) along with its simplified form for incompressible plasmas, in an untwisted tube, as well as that for waves running along an incompressible twisted jet surrounded by also incompressible ambient coronal plasma, Eq.~(\ref{eq:twdispeq}).  Numerical calculations show that for the physical parameters of the observed EUV jet and its environment the kink, $m = 1$, mode become unstable for jet speeds in the range of $112$--$114.8$~km\,s$^{-1}$---speeds that are within observationally registered speed of $150$~km\,s$^{-1}$.  The question that immediately arises is how any change in physical parameters of the jet or its environment will influence the condition for KH instability occurrence.  If we decrease, for instance, the electron density of the ambient coronal plasma to $5 \times 10^{10}$~cm$^{-3}$, making the density contrast $\eta$ equal to $0.5$ (that corresponds to a $2$-times denser jet than its environment), the pressure balance equation cannot be satisfied for $B_\mathrm{e} = 7$~G.  If we still want to keep that value, one must reduce the temperature inside the jet---$T_\mathrm{i} = 900\,000$~K is an appropriate choice and we have a new value for the reference Alfv\'en speed, notably $v_\mathrm{Ai} = 63.09$~km\,s$^{-1}$.  Performing all the computations with this new $\eta = 0.5$ and $b = 0.765$ (for the untwisted tube we need also of $\tilde{\beta}_\mathrm{i} = 3.11$ and $\tilde{\beta}_\mathrm{e} = 5.91$) we obtain that the kink, $m = 1$, mode (in both configurations) and harmonic MHD modes in a twisted magnetic tube with $\varepsilon = 0.025$ can become unstable at flow speeds of $140.3$~km\,s$^{-1}$ (at $M_\mathrm{A}^\mathrm{cr} = 2.224058$) for the compressible untwisted tube, and $147.5$, $143$, $141.8$, and $141.6$~km\,s$^{-1}$ correspondingly for the $m = 1,2,3,4$ modes in the incompressible tube with also incompressible environment.  As seen, all these critical speeds are still accessible for our EUV jet.  On the other hand, if we choose the background magnetic field to be equal to $10$~G (at $\eta = 0.589$), that would imply that the reference Alfv\'en speed will increase to $68.457$~km\,s$^{-1}$, the critical jet speed for triggering a KH instability of the kink, $m = 1$, mode in an untwisted jet at $M_\mathrm{A}^\mathrm{cr} = 2.31995$ will become equal to $158.8$~km\,s$^{-1}$---a speed that is inaccessible for our EUV jet.  The critical velocities for all modes in the twisted tube turn out to be even higher: they are equal to $164.7$, $162$, $161.8$, and $161.5$~km\,s$^{-1}$ for the $m = 1,2,3,4$ MHD modes, respectively.  This conclusion confirms the famous Chandrasekhar's statement \citep{chand61} that a longitudinal magnetic field can, in general, stabilize the jet against the KH instability.

In their recent paper, \cite{aja} exploring the KH instability of the kink ($m =1$) mode in Type II spicules performed an interesting plasma and magnetic field parameter study for the instability onset.  In the beginning, as an illustration, they investigated the stability/instability status of kink waves traveling along a spicule with a density contrast $\eta = 0.1$ (too low for these kind of chromospheric jets) and a ratio of axial magnetic fields, $b = 0.25$.  They modeled the spicule as a cool plasma surrounded by incompressible magnetized plasma.  Dispersion curves and growth rates of unstable kink mode for various Alfv\'en Mach numbers (that is, jet velocities) plotted in their Fig.~1 in many aspects are similar to our Fig.~\ref{fig:fig2}.  It is necessary, however, to emphasize that the marginal growth rate curve should be refined, notably it is naturally to consider as a marginal growth rate curve the one that touches (at levels in the range of $0.003$--$0.005$) the $k_z a$-axis---notably that curve defines the actual threshold Alfv\'en Mach number.  (The same criticism applies to Fig.~5 in \citealp{zhel12a}.)  We have also spotted that the growth rate curves in Fig.~1 in \citealp{aja} corresponding to Alfv\'en Mach numbers equal to $3.93$ and $3.97$ are not
\begin{figure}[!ht]
  \centering
\subfigure{\includegraphics[width = 3.3in]{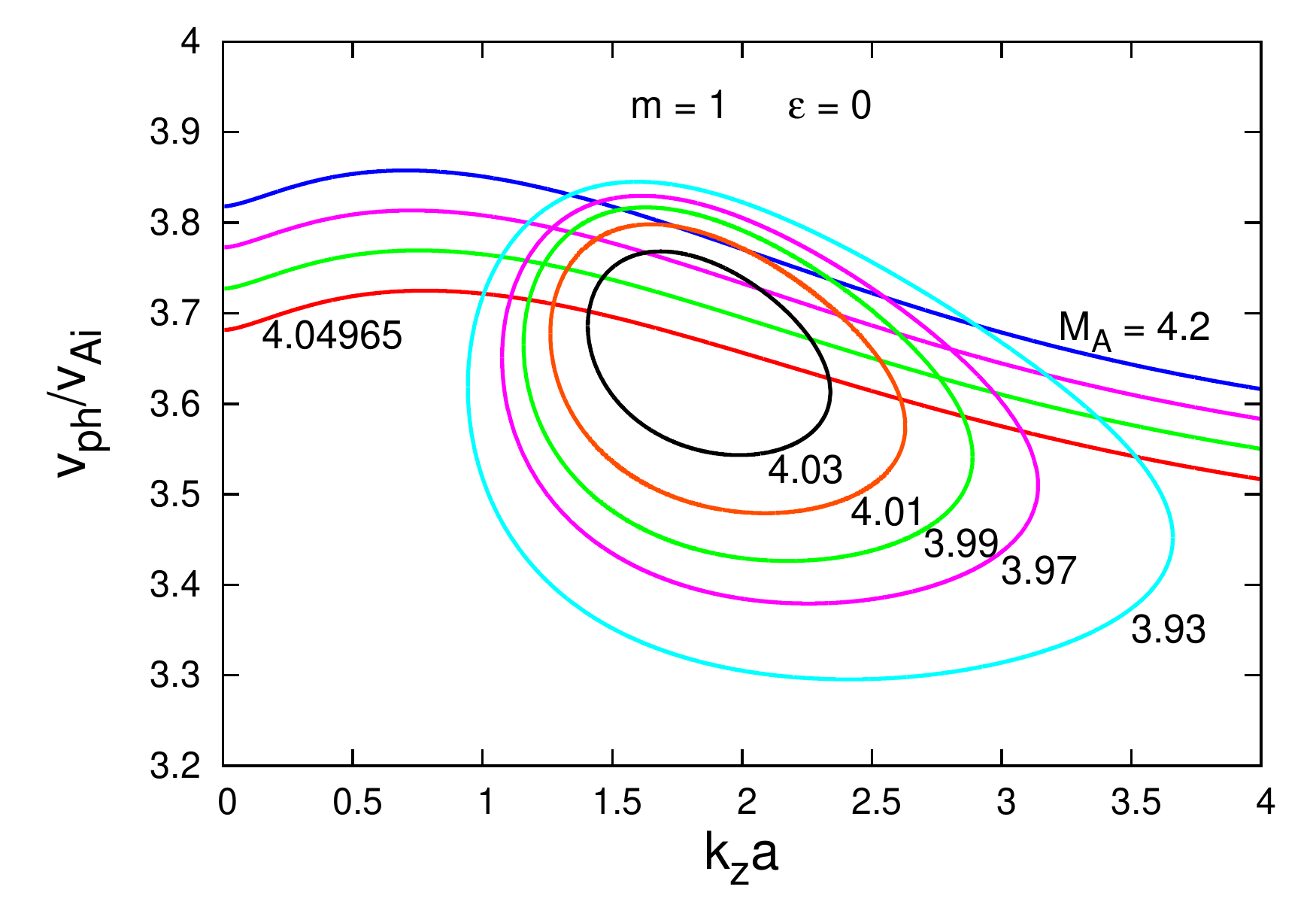}} \\
\subfigure{\includegraphics[width = 3.3in]{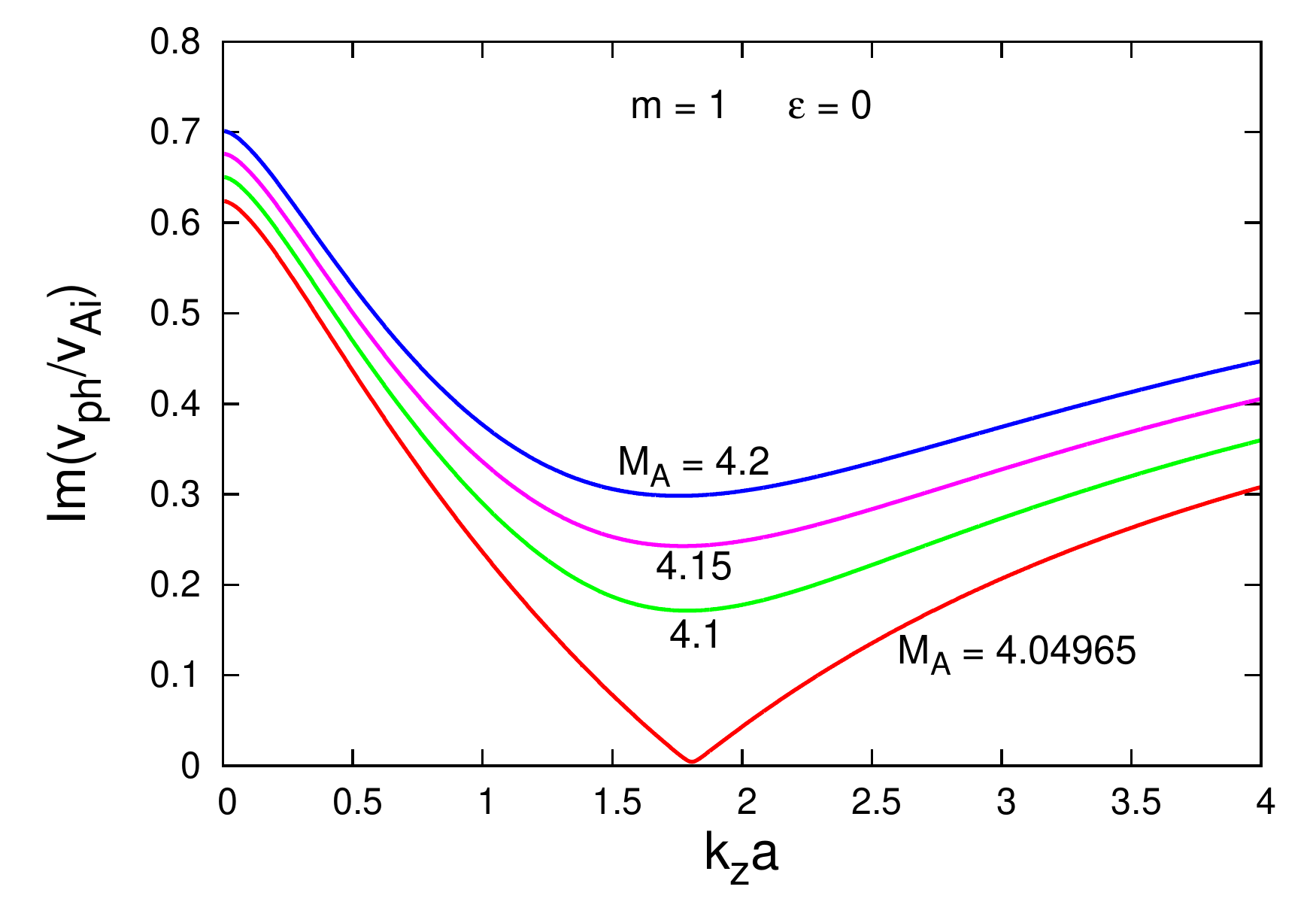}}
  \caption{(\emph{Top panel}) Dispersion curves of unstable kink ($m = 1$) MHD mode propagating on a moving untwisted flux tube of cool plasma surrounded by incompressible magnetized plasma at $\eta = 0.1$ and $b = 0.25$ for four values of the Alfv\'en Mach number $M_\mathrm{A} = 4.04965$, $4.1$, $4.15$, and $4.2$.  Closed dispersion curves correspond to stable kink wave propagation.  (\emph{Bottom panel})  The normalized growth rates for the same values of $M_\mathrm{A}$.  Red curves in both plots correspond to the onset of KH instability.}
  \label{fig:fig8}
\end{figure}
adequate---our computations with their input parameters show that the dispersion curves associated with these two numbers depict stable kink mode propagation (see the top panel in Fig.~\ref{fig:fig8}).  The marginal growth rate curve was obtained at $M_\mathrm{A}^\mathrm{cr} = 4.04965$ that yields with a reference Alfv\'en speed of $40$~km\,s$^{-1}$ a critical jet speed of $162$~km\,s$^{-1}$, being higher that the maximal Type II spicules' speed of $150$~km\,s$^{-1}$.  On the other hand, their curves plotted in Fig.~4 in \citealp{aja} seem to be correct.  One can see, that according to the
curve corresponding to $\eta = 0.1$ for $b = 0.25$ the critical Alfv\'en Mach number, $M_\mathrm{A}^\mathrm{cr}$, must be a little higher than $4$, in fine agreement with our computations.  We have checked also that with $\eta = 0.01$ the critical Alfv\'en Mach numbers corresponding to $b$ equal to $0$, $0.36$, and $1$ are $11.991254$, $12.517196$ (a refined value from Fig.~5 in \citealp{zhel12a}), and $15.64737$, respectively, and all they are in good agreement with the highest curve in Fig.~4 in \citealp{aja}.  Another interesting observation is the circumstance that the minima of all computed by us `v'-shaped (see the red curve in the bottom panel of Fig.~\ref{fig:fig8}) growth rate curves were grouped around $k_z a = 1.6$, more specifically $1.609$ at $b = 0$, $1.611$ for $b = 0.36$, and $1.623$ with $b = 1$.

\cite{and} were the first to derive a criterion for predicting the rising of KH instability in untwisted magnetic flux tubes (in the limit of cold plasma, $\beta_\mathrm{i} = 0$), notably it (instability) occurs if (in our notation)
\begin{equation}
\label{eq:andries}
    M_\mathrm{A} > 1 + b/\sqrt{\eta}.
\end{equation}
In their paper, \cite{aja}, on using asymptotic presentations of modified Bessel functions, claim that the KH instability will emerge if
\begin{equation}
\label{eq:hossein}
    M_\mathrm{A} > \left[ \frac{\eta + 1}{\eta}\left( b^2 + 1 \right) \right]^{1/2}.
\end{equation}
We note that this criterion generalize the criterion of \cite{holz} applicable for untwisted tube with non-magnetic environment, that is, $B_\mathrm{e} = 0$ (see Eq.~(5) in \citealp{holz}).  It is intriguing to see how accurately these two criteria predict the threshold/critical Alfv\'en Mach number for various studies of KH instability in untwisted moving flux tubes over recent years.  In Table \ref{tbl:table1} we present the input data ($\eta$ and $b$) for exploring the KH instability in spicules \citep{zhel12a,zhel13,aja}, X-ray jets \citep{vash,zhel13}, photospheric tubes \citep{zhel13}, and EUV jets (present paper).  It is clearly seen that the Andries \& Goossens criterion (\ref{eq:andries}) provides a fairly good prediction only in the case when the ratio of axial magnetic
\begin{table*}
\small
\caption{Predicted and computed critical Alfv\'en Mach numbers\label{tbl:table1}}
\begin{tabular}{@{}lcccc@{}}
\tableline
Paper & Input data & $M_\mathrm{A}$ from Eq.~(\ref{eq:andries}) & $M_\mathrm{A}$ from Eq.~(\ref{eq:hossein}) & $M_\mathrm{A}$ computed \\
\tableline
\cite{vash}    & $\eta = 0.13$, $b \cong 1.04$ & $3.871$ & $4.243$ & $4.47$ \\
\cite{zhel12a} & $\eta = 0.01$, $b = 0.36$ & $4.6$ & $10.681$ & $12.517196$ \\
\cite{zhel13} & $\eta = 0.02$, $b = 0.35$  & $3.475$ & $7.566$ & $8.85925$ \\
\cite{zhel13} & $\eta =2.00$, $b = 0.07$ & $1.049$ & $1.228$ & $1.2523295$ \\
\cite{zhel13} & $\eta = 0.13$, $b \cong 1.04$ & $3.871$ & $4.243$ & $4.2485$ \\
\cite{aja} & $\eta = 0.10$, $b = 0.25$ & $1.791$ & $3.419$ & $4.04965$ \\
This paper & $\eta \cong 0.60$, $b = 1.01$  & $2.312$ & $2.329$ & $2.354327$ \\
This paper & $\eta = 0.50$, $b = 0.76$ & $1.823$ & $2.18$ & $2.224058$ \\
This paper & $\eta \cong 0.60$, $b = 1.00$ & $2.311$ & $2.320$ & $2.31995$ \\
\tableline
\end{tabular}
\end{table*}
fields, $b$, is of the order of $1$.  The new criterion (\ref{eq:hossein}) of Ajabshirizadeh et al.\ is in any way superior to that of Andries \& Goossens---it yields values rather close to the computed ones.  Nevertheless, we should not forget that the obtaining of a correct threshold/critical Alfv\'en Mach number, $M_\mathrm{A}^\mathrm{cr}$, can be done by carefully studying the evolution of the pair of kink-speed dispersion curves with gradually increasing $M_\mathrm{A}$ to find the marginal growth rate curve---this procedure sometimes requires many takes during the computation.  The final criterion, of course, is the product of $M_\mathrm{A}^\mathrm{cr}$ and the reference Alfv\'en speed that defines the critical flow velocity for the KH instability onset.

An instability criterion for the case of moving twisted magnetic flux tubes was obtained by analytically solving dispersion equation (\ref{eq:twdispeq}) in long wavelength approximation \citep{zaq14} and it reads in our notation as
\begin{equation}
\label{eq:temury}
    |m|M_\mathrm{A}^2 > \left( 1 + \frac{1}{\eta} \right)\left( |m|b_\mathrm{twist}^2 + 1 \right).
\end{equation}
This inequality is a generalization of an instability criterion of twisted magnetic tubes with non-magnetic environment that was derived by \cite{zaq10}.  It is easy to observe that Eq.~(\ref{eq:temury}) contains the case of an untwisted flux tube, in which case $|m| = 1$ and $b_\mathrm{twist} \equiv b$---then Eq.~(\ref{eq:temury}) coincides with Eq.~(\ref{eq:hossein}).  In other words, \cite{aja} reinvented in their own way an already published result.

The threshold Alfv\'en Mach numbers for the kink ($m = 1$) mode and higher $m$ modes ($m = 2$--$4$) in our EUV jet modeled as a moving twisted magnetic tube according to Eq.~(\ref{eq:temury}) are equal to $2.329$ for the kink mode and to $2.022$, $1.909$, and $1.849$ for the higher $m$ modes, respectively.  In this case of twisted tube the discrepancy between predicted and computed values of the threshold Alfv\'en Mach number is generally larger than in the case of untwisted tube---now we have $2.329$ vs.\ $2.4068$ for the kink ($m = 1$) mode, $2.367$ vs.\ $2.022$ for the fluting-like ($m = 2$) mode, $2.365$ vs.\ $1.909$ for the $m = 3$ harmonic, and $2.363$ vs.\ $1.849$ for the $m = 4$ mode, respectively.  We emphasize on the big difference between the shapes of the growth rate curves of the kink mode and those of the higher $m$ modes: while the marginal growth rate curve of kink mode has the typical `v' shape observed in untwisted tubes (compare red curves in the bottom panels of Figs.~\ref{fig:fig2} and \ref{fig:fig4}), the marginal growth rate curves of higher $m$ modes are more or less similar in form, too, but they have a distinctive property, namely they start at some critical $k_z a$ (or wavelength) numbers gradually shifted to the right on the $k_z a$-axis (look at the plots in the top panels of Figs.~\ref{fig:fig5}--\ref{fig:fig7}). For all the MHD modes excited in twisted tubes the wave growth rates are of the order of a few to dozens inverse milliseconds depending on the magnitude of the KH unstable wavelength.

\section{Conclusion}
\label{sec:concl}
In this paper, we carried out an exploration of the possibility for the development of KH instability in an EUV solar jet observed in an active coronal region.  We modeled the jet as a moving magnetic flux tube in two geometries, notably first as a cylindrical untwisted tube and after that as a weakly uniformly twisted magnetic tube.  The twist of the magnetic field $\vec{B}_\mathrm{i}$ inside the tube is characterized by the twist parameter $\varepsilon$, defined as the ratio of azimuthal to axial components of $\vec{B}_\mathrm{i}$ evaluated at the boundary $r = a$ of the tube.  The ambient coronal magnetic field, $\vec{B}_\mathrm{e}$, is supposed to be homogeneous, as are the electron densities $\rho_\mathrm{i}$ and $\rho_\mathrm{e}$ in both media, the jet and its environment.  The jet flow velocity, $\vec{v}_0$, being constant in radial direction, exhibits a jump at the tube boundary that may result in emergence of the KH instability under a specific critical value of the jet speed.  The primary goal of our study was to establish whether the EUV jet under consideration (whose flowing velocity is $150$~km\,s$^{-1}$) can become unstable against KH instability.  Our, largely numerical, analysis is based on solving the dispersion relations of various MHD modes propagating along the moving tube (be it untwisted or twisted).  We have used well-known wave dispersion relations derived from the governing equations of ideal magnetohydrodynamics (see, e.g., \citealp{frei}) in two approximations: of compressible plasmas for the untwisted magnetic tube and in the incompressible limit for the moving twisted tube.  In searching for the instability conditions, we assumed that the wave frequency, $\omega$, is complex, and by numerically solving the appropriate dispersion relations in complex variables by varying the flow velocity (that is, the Alfv\'en Mach number, $M_\mathrm{A} = v_0/v_\mathrm{Ai}$, in the dimensionless form of wave dispersion equations~(\ref{eq:dispeq}) and (\ref{eq:twdispeq})) we have found so called marginally dispersion and growth rates curves that determine the instability onset.  The threshold/critical Alfv\'en Mach number, associated with the marginal curves, yields the seeking critical velocity, $v_0^\mathrm{cr} = M_\mathrm{A}^\mathrm{cr} v_\mathrm{Ai}$.  We have obtained that all studied MHD modes are subject to the KH instability at speeds below the observationally registered speed of $150$~km\,s$^{-1}$.  The lowest critical velocity of $112$~km\,s$^{-1}$ is that of the kink, $m = 1$, mode traveling on an untwisted flux tube of compressible plasma---that velocity of the same mode, but for incompressible plasma both in untwisted and weakly twisted magnetic tube ($\varepsilon = 0.025$) turns out to be higher (${=}114.8$~km\,s$^{-1}$ in the untwisted tube and $114.4$~km\,s$^{-1}$ in twisted one).  In a twisted moving tube the KH instability of higher $m$ modes, $m = 2$--$4$, MHD modes may occur if the EUV jet speed is equal to ${\cong}112.4$~km\,s$^{-1}$.  Obviously, there is no big differences between all these critical velocities irrespective of the magnetic tube geometry and excited MHD modes.

KH unstable modes are generally surface waves but their type crucially depends on the ordering of Alfv\'en and sound speeds in the jet and its environment \citep{call}.  The kink ($m = 1$) mode propagating on a moving untwisted magnetic flux tube of compressible plasma (the model of our EUV jet) being initially, for $0 \leqslant M_\mathrm{A} < M_\mathrm{A}^\mathrm{cr}$, a pure surface wave becomes a leaky wave when $M_\mathrm{A} \geqslant M_\mathrm{A}^\mathrm{cr}$.  By contrast, in the incompressible limit, the unstable kink modes both in untwisted and twisted flux tubes are non-leaky surface waves.  This observation shows once again that the plasma compressibility plays an important role in studying waves and instabilities in bounded magnetized plasmas.

A hint in looking for the instability conditions for the excitation of KH instability of various MHD modes in moving tubes, one may use the instability criterion (\ref{eq:temury}) that we present here in its original form (see Eq.~(30) in \citealp{zaq14})
\[
    |m|M_\mathrm{A}^2 > \left( 1 + \frac{\rho_\mathrm{i}}{\rho_\mathrm{e}} \right) \left( |m|\frac{B_\mathrm{e}^2}{B_{\mathrm{i}z}^2} + 1 \right).
\]

Recall that in the case of untwisted magnetic flux tube $|m| = 1$ and $B_{\mathrm{i}z} = B_\mathrm{i}$.  Although this inequality provides sometimes rather good predicting values of $M_\mathrm{A}^\mathrm{cr}$, the best way to obtain the actual threshold value is to carefully study the evolution of wave dispersion curves with gradually increasing (or, better, decreasing) $M_\mathrm{A}$ in the neighborhood of its predicted critical value.

When the jet is modeled as an untwisted moving flux tube, it is preferable to solve dispersion equation (\ref{eq:dispeq}) for compressible plasma because it yields the lowest $M_\mathrm{A}^\mathrm{cr}$.  KH vortices occurring during the nonlinear evolution of the KH instability can be considered as one of the important sources of MHD wave turbulence in solar jets \citep{vanball,asgari}.  EUV jets, with their typical physical characteristics, and, most importantly, their flow speeds, can be considered as a bridge between, say, the surges and Type II spicules from one side, and the soft X-rays jets and CMEs from the other side.  In any case, the EUV jets, especially those emerging from coronal holes, might contribute to supplying energy and material in the solar wind.  In that respect, their study using more sophisticated models considering inhomogeneity, viscosity, gravity, flow velocity and/or magnetic field
shear, and nonlinearity effects can significantly improve
our understanding of the jet-like phenomena in the solar dynamic atmosphere.

%
\acknowledgments
This work was supported by
\vspace{198mm}
the Bulgarian Science Fund and the Department of Science \& Technology, Government of India Fund under Indo-Bulgarian bilateral project CSTC/INDIA 01/7, /Int/Bulgaria/P-2/12.  The authors would like to thank the anonymous reviewer for her/his helpful and constructive comments that greatly contributed to improving the final version of the manuscript.  We are also indebted to Snezhana Yordanova for drawing one figure.


%

\end{document}